\documentclass[
    10pt,
    onecolumn,
    secnumarabic,
    superscriptaddress,
    nobibnotes,
    nofootinbib,
    showkeys,
    aps,
    prd
]{revtex4-2}

\usepackage[T1]{fontenc}       % Better font encoding
\usepackage{amsmath, amssymb}  % Math symbols and environments
\usepackage{amsfonts}          % Extra math fonts
\usepackage{graphicx}          % Figures and graphics
\usepackage{bm}                % Bold math symbols
\usepackage{xcolor}            % Color support
\usepackage{comment}           % Commenting out blocks
\usepackage{hyperref}          % Hyperlinks
\usepackage{caption}
\usepackage{subcaption}
\providecommand{\U}[1]{\protect\rule{.1in}{.1in}}

\usepackage{tocloft} % For customizing table of contents, figures, and tables

\providecommand{\U}[1]{\protect\rule{.1in}{.1in}}
\newcommand{\be}{\begin{equation}}
\newcommand{\ee}{\end{equation}}

\newcommand{\mincir}{\raise
-3.truept\hbox{\rlap{\hbox{$\sim$}}\raise4.truept\hbox{$<$}\ }}
\newcommand{\magcir}{\raise
-3.truept\hbox{\rlap{\hbox{$\sim$}}\raise4.truept\hbox{$>$}\ }}

\begin{document}
\title{Compartmentalization in the Dark Sector of the Universe after DESI DR2 BAO data}
\author{Marcel van der Westhuizen}
\email{marcelvdw007@gmail.com }
\affiliation{Centre for Space Research, North-West University, Potchefstroom 2520, South Africa}
\author{David Figueruelo}
\email{david.figueruelo@ehu.eus}

\affiliation{Centre for Space Research, North-West University, Potchefstroom 2520, South Africa}
\affiliation{Department of Theoretical Physics, University of the Basque Country UPV/EHU, P.O. Box 644, 48080 Bilbao, Spain}
\author{Rethabile Thubisi}
\email{feliciarethabile@gmail.com}
\affiliation{Centre for Space Research, North-West University, Potchefstroom 2520, South Africa}

\author{Shambel Sahlu}
\email{shambel.sahlu@nithecs.ac.za}
\affiliation{Centre for Space Research, North-West University, Potchefstroom 2520, South Africa}
\affiliation{Department of Physics, Wolkite University, Wolkite, Ethiopia}
\author{Amare Abebe}
\email{Amare.Abebe@nithecs.ac.za}
\affiliation{Centre for Space Research, North-West University, Potchefstroom 2520, South Africa}
\affiliation{National Institute for Theoretical and Computational Sciences (NITheCS), South Africa}
\author{Andronikos Paliathanasis}
\email{anpaliat@phys.uoa.gr}
\affiliation{Centre for Space Research, North-West University, Potchefstroom 2520, South Africa}
\affiliation{National Institute for Theoretical and Computational Sciences (NITheCS), South Africa}
\affiliation{School for Data Science and Computational Thinking and Department of
Mathematical Sciences, Stellenbosch University, Stellenbosch, 7602, South Africa}
\affiliation{Departamento de Matem\`{a}ticas, Universidad Cat\`{o}lica del Norte, Avda. Angamos 0610, Casilla 1280 Antofagasta, Chile}

\begin{abstract}
We study a non-linear interaction between the dark matter and dark energy components of the universe. Within a spatially flat FLRW geometry, dark matter is described by a dust fluid and dark energy by an ideal gas with a constant equation-of-state parameter, allowing for energy transfer between the two sectors. The resulting effective fluid leads to a unified dynamical dark energy model in which the Hubble function admits an analytic expression. We constrain this model using the DESI DR2 Baryon Acoustic Oscillation (BAO) measurements and the Pantheon+ Type Ia supernovae (SNe Ia) compilation. For the combined dataset, the interacting model provides a better fit than $\Lambda$CDM, with $\chi^2_{\rm Interaction}-\chi^2_{\Lambda{\rm CDM}}=-5$. Using the Akaike Information Criterion (AIC), we find $\text{AIC}_{\rm Interaction}-\text{AIC}_{\Lambda{\rm CDM}}=-1$, indicating that the interacting model is weakly preferred, though the difference is not statistically significant and disappears when considering BAO data alone. Finally, when comparing the interaction with the $w_0w_a$CDM (CPL) model, we find a small preference for the interaction in the BAO-only case, with $\text{AIC}_{\rm Interaction}-\text{AIC}_{w_0w_a}=-1.5$.
\end{abstract}
\keywords{cosmology, interactions, cosmological constraints, dark energy}\date{\today}
\maketitle

\section{Introduction}
Dynamical dark energy models have recently regained attention in the literature after the release of the DESI DR2 collaboration on the Baryonic Acoustic Oscillations (BAO) \cite{DESI:2025zgx,DESI:2025fii,DESI:2025qqy,DESI:2025wyn}. Specifically, it has been shown that the $w_{0}w_{a}$CDM parametric models are supported by the cosmological data, to the extent that it is favored over the concordance $\Lambda$CDM model. This is in agreement with last year's
conclusions for DESI DR1 \cite{DESI:2024uvr,DESI:2024mwx} and by a series of
independent studies \cite{Park:2024vrw,Park:2024pew,Park:2025azv} on other
datasets.  
In the literature, various cosmological theories have been tested as dynamical
dark energy models by using the DESI DR2 BAO data, see for instance
\cite{Ormondroyd:2025iaf,Yang:2025mws,Li:2025cxn,Paliathanasis:2025hjw,Tyagi:2025zov,You:2025uon, Santos:2025wiv,Alfano:2025gie,Chaussidon:2025npr,Anchordoqui:2025fgz,Ye:2025ulq,Silva:2025hxw,Wolf:2025jed,Paliathanasis:2025dcr,Paliathanasis:2025cuc,Nagpal:2025omq,Chen:2025mlf,Wang:2025bkk,Kumar:2025etf,Luciano:2025hjn,Dinda:2025svh, choudhury2025cosmologyextendedparameterspace}. Between the different gravitational models with dynamical scalar fields, or modifications of the Einstein-Hilbert Action, cosmological interactions can provide a mechanism for a dynamical interacting model from a simple physical starting point.  

Cosmological models with interaction in the dark sector of the universe, that is, dark matter-dark energy interaction, have long been studied in the literature (see Ref. \cite{Wang:2024vmw} for a review). Interacting models were initially proposed for their potential to address the coincidence problem \cite{Amendola_2000, Zimdahl_2001,Chimento_2003, Farrar_2004, Wang_2004, Olivares_2006}, which is related to the observation that the ratio of dark matter to dark energy is unity at present, while this ratio is predicted to differ by many orders of magnitude in both the past and future.
Although those interacting models mainly emerge from a phenomenological approach, from a theoretical standpoint, and specifically within the framework of particle physics, it is feasible for two matter fields, such as dark matter and dark energy, to interact. Moreover, one significant finding is that permitting such an interaction can shift the dark energy equation of state from the quintessence regime to the phantom regime without the appearance of ghosts or big rip singularities \cite{Curbelo:2005dh}. 
Parallel to that theoretical motivation, in the last few years and thanks to the enormous precision achieved by the observational probes, models which describe interactions have become popular for their potential to alleviate the well-known cosmological tensions~\cite{Abdalla:2022yfr,Perivolaropoulos:2021jda}. Particularly pressing are the so-called $H_0$ crisis and the $\sigma_8$ tension. Here, interacting models have shown variable levels of success, depending on the choice of interaction kernel and data sets used \cite{CosmoVerse:2025txj}. A specific class of interacting models was recently studied in \cite{Shah:2025ayl,Pan:2025qwy} by using the DESI DR2 BAO data, while previous studies of other interacting models using DESI DR1 BAO data can be found in \cite{Li:2024qso, Li:2025owk}.

In this work, we investigate the cosmological viability of an interacting model which leads to compartmentalization in the dark sector of the universe. Specifically, we consider the non-linear interaction \cite{Koshelev:2010umw}
\begin{equation}
Q\left(  t\right)  =Q_{0}(t)\rho_{\rm{dm}}\rho_{\rm{de}} \;,\quad \text{where} \quad Q_{0}\left(  t\right)
=\frac{\delta}{H}. \label{in.01}%
\end{equation}

We have analyzed the late-time case where we can make the approximation $3H^2=\rho_{\rm{dm}}+\rho_{\rm{de}}$ 
which leads to the equivalent interaction kernel: 
\begin{equation} \label{Q.dmde}
Q=3H\delta \left( \frac{\rho_{\text{dm}}\rho_{\text{de}} } {\rho_{\text{dm}}+\rho_{\text{de}}} \right).
\end{equation}
This interacting model is known to be free from early instabilities, as demonstrated in Ref.~\cite{Koshelev:2010umw}. Moreover, the same interaction term was examined in Ref.~\cite{Paliathanasis:2024abl}, where it was shown to describe compartmentalization within the dark sector of the universe. The full background dynamics of this interaction has recently been studied in \cite{vanderWesthuizen:2025II}.

Derivations of the perturbation equations and a full study of large-scale instabilities fall outside the scope of this work, but can be found for this exact kernel in Ref.~\cite{Li_2014}, in which they demonstrated that cosmological perturbations remain stable during the entire expansion history, provided that $\delta<0$ when $\omega>-1$. This is in agreement with the doom factor analysis of Ref.~\cite{M.B.Gavela_2009, Honorez_2010}, which defines two stable regimes: (i) energy flow from dark matter to dark energy ($\delta<0$ combined with $\omega>-1$), and (ii) energy flow from dark energy to dark matter ($\delta>0$ combined with $\omega<-1$). 

In addition to stability, the interaction has implications for structure growth and early-universe observables. At the level of the CMB, the coupling modifies the effective expansion rate and can shift the sound horizon at recombination, thereby altering the position of acoustic peaks. For large-scale structure, the direction of energy transfer determines the growth history of matter perturbations. When energy flows from dark energy to dark matter ($\delta>0$), the effective matter density is reduced in the past, leading to a suppression of the growth of perturbations and a lower matter power spectrum; conversely, when energy flows from dark matter to dark energy ($\delta<0$), the effective matter density is enhanced in the past, resulting in stronger clustering and an enhanced matter power spectrum \cite{Lucca:2021dxo, vanderWesthuizen:2023hcl}. While we do not perform a perturbation-level analysis here, these qualitative considerations show that the main deviations from $\Lambda$CDM are expected at late times, whereas the model can remain consistent with early-universe observations under the above stability conditions.

Previous observational constraints of this model are discussed in Refs.~\cite{von_Marttens_2019, Arevalo:2011hh, Aljaf_2021}, while its potential to address the $H_{0}$ and $\sigma_{8}$ tensions is explored in~\cite{Yang_2023}. This interaction has also been analyzed within holographic dark energy models \cite{Feng_2016}, and the interaction kernel exhibits behaviour similar to Chaplygin gas models~\cite{Zhang_2006, Li_2014}, particularly in relation to statefinder diagnostics~\cite{carrasco2023discriminatinginteractingdarkenergy}.
In addition to the aforementioned properties, a notable feature of this model is its ability to avoid the negative energy densities that commonly arise in interacting dark sector scenarios. As previously demonstrated in Ref.~\cite{vanderWesthuizen:2025I, vanderWesthuizen:2025II}, the energy transfer mechanism naturally ceases when either the dark matter or dark energy density approaches zero. This prevents the energy crossing into unphysical negative values, both in the past and future. 
Another important characteristic of this interacting model is that it leads to an
analytic solution for the field equations. This property yields an analytic expression for the Hubble function and the energy density equation that allows us to perform analyses of the data and determine the best fit values for the free parameters.

The structure of the paper is as follows. Section \ref{section2} presents the cosmological model of our consideration, which
describes interactions within the dark sector. Section \ref{section3} includes the main results of this study, where we compare our cosmological theory with the recent cosmological observations. In particular, we use the Pantheon+ data and the BAO from the DESI DR2 collaboration, and compare our results to both the $\Lambda$CDM model and $w_0w_a$CDM model. Finally, in Section \ref{section4}, we draw our conclusions. 

\section{Compartmentalization in the Dark Sector}\label{section2}

In this section, we present the relevant equations that govern the
dynamics of the model.  

We assume general relativity in a homogeneous and isotropic universe, described by the 
Friedmann – Lemaître – Robertson – Walker (FLRW) metric with no spatial curvature
defined as
\begin{equation}
ds^{2}=-\mbox{\rm d}t^{2}+a^{2}(t)\mbox{\rm d}\Vec{x}^{2}\;,\label{metric1}%
\end{equation}
where $t$ is the cosmic time, $\Vec{x}$ are the comoving spatial coordinates,
and the function $a(t)$, the scale factor, describes the radius of a
three-dimensional space, while the expansion rate for the comoving observer
$u^{\mu}=\delta_{t}^{\mu}$ is $\theta=3H$, where $H$ is the Hubble function
defined as $H=\frac{\dot{a}}{a}$, and $\dot{}$ denotes differentiation with respect to cosmic time. 
We will also assume that each of the universe components can be described by a stress-energy tensor of perfect fluids as
\begin{equation}
T_{\textrm{i}}^{\mu\nu}=(\rho_{\text{i}}+p_{\textrm{i}})u_{\textrm{i}}^{\mu
}u_{\textrm{i}}^{\nu}+p_{\textrm{i}}g^{\mu\nu}\;,
\end{equation}
where $\rho_{\textrm{i}}$, $p_{\textrm{i}}$ and $u_{\textrm{i}}^{\mu}$ represent the
energy density, pressure, and 4-velocity of the ${\textrm{i}}$-component of the
Universe. In our case, we focus on the epoch dominated by pressureless
matter, namely baryons and dark matter ($T^m_{\mu\nu}$), and dark energy ($T^{de}_{\mu\nu}$), which are described
by
\begin{align}
T_{\mu\nu}^{\textrm{m}} &  =\rho_{m}u_{\mu}u_{\nu}\;,\\
T_{\mu\nu}^{\textrm{de}} &  =\left(  \rho_{\textrm{de}}+p_{\textrm{de}}\right)
u_{\mu}u_{\nu}+p_{\textrm{de}}g_{\mu\nu}\;.
\end{align}
Given the previous considerations, the gravitational field equations read as
\begin{equation}
R_{\mu\nu}-\frac{R}{2}g_{\mu\nu}=T_{\mu\nu}^{\textrm{m}}+T_{\mu\nu}^{\textrm{de}%
}\;,
\end{equation}
while from the Bianchi identity we determine the conservation
\begin{equation}
\left(  T_{\mu\nu}^{\textrm{m}}g^{\nu\kappa}+T_{\mu\nu}^{\textrm{de}}g^{\nu\kappa
}\right)  _{;\kappa}=0\;,
\end{equation}
since the total stress energy tensor is conserved but the individual species are not. This conservation of the total dark sector can be encapsulated in a modification of the conservation equations for dark matter and dark energy given by \eqref{conservationEqdm} and \eqref{conservationEqde}.
\begin{align} \
\dot{\rho}_{\textrm{dm}}+3H\rho_{\textrm{dm}} &  =Q(t)\;,\label{conservationEqdm} \\\
\dot{\rho}_{\textrm{de}}+3H\left(  \rho_{\textrm{de}}+p_{\textrm{de}}\right)   &
=-Q(t)\;, \label{conservationEqde}
\end{align}
where $Q\left(  t\right)  $ introduces interaction between the two fluids and
represents the definition of the energy transfer as presented in Eq.  \eqref{in.01}. 
It follows easily that for $Q\left(  t\right)  >0$, there is energy transfer from dark energy to dark matter, while for $Q\left(  t\right)  <0$, there is energy transfer from dark matter to dark energy. \\
Finally, we consider dark energy to satisfy the
closure relation between its energy density and pressure as $p_{\textrm{de}}=w\rho_{\textrm{de}}$, where $w$ is the equation of state which we assume to be
constant. For the particular choice of equation~\eqref{in.01}, analytical solutions can
be derived for the time evolution of the dark matter and dark energy
interacting components as found in \cite{Arevalo:2011hh, Li_2014, vanderWesthuizen:2025II}, which read as :
\begin{equation}%
\rho_{\textrm{dm}}    =\rho_{\textrm{(dm,0)}}a^{-3(1-\delta
)}\left[  \frac{1+ \left( \frac{\rho_{{\textrm{(dm,0)}}}}{\rho_{{\textrm{(de,0)}}%
}} \right)  a^{3\left[  w+\delta\right]  }}{1+ \frac{\rho_{{\textrm{(dm,0)}}}%
}{\rho_{{\textrm{(de,0)}}}} } \right]  ^{-\frac{\delta}{(w+\delta)}} \;, \label{AnalyticalSol}
\end{equation}
and
\begin{equation}
\rho_{\textrm{de}}    =\rho_{\textrm{(de,0)}}a^{-3(1+w)}\left[  \frac{1+ \left(
\frac{\rho_{{\textrm{(dm,0)}}}}{\rho_{{\textrm{(de,0)}}}} \right)  a^{3\left[
w+\delta\right]  }}{1+ \frac{\rho_{{\textrm{(dm,0)}}}}{\rho_{{\textrm{(de,0)}}}} }
\right]  ^{-\frac{\delta}{(w+\delta)}} \;,\label{eq01}%
\end{equation}
where $\delta$ is the parameter controlling the interaction as previously
explained. It should be noted that we require $\delta\ne-w$ in order to
avoid undefined solutions for \eqref{AnalyticalSol} and \eqref{eq01}. Thus, the  normalized Hubble function   \(\left(  \frac{H_{C}\left(  z\right)  }{H_{0}}\right)  ^{2}\equiv E_{C} %
^{2}\left(  z\right)\), for a universe dominated by matter and dark energy is
\begin{equation}
 E_{C}%
^{2}\left(  z\right) =\left(  \hat{\Omega}_{{\textrm{(dm,0)}}}\left(
1+z\right)  ^{3\left(  1-\delta\right)  }+\Omega_{{\textrm{(de,0)}}}\left(  1+z\right)  ^{3\left(  1+w\right)  }\right)  \left(  \frac{1+\left(
\frac{\hat{\Omega}_{{\textrm{(dm,0)}}}}{\Omega_{{\textrm{(de,0)}}}}\right)  \left(  1+z\right)
^{-3\left(  w+\delta\right)  }}{1+\left(  \frac{\hat{\Omega}_{{\textrm{(dm,0)}}}}
{\Omega_{{\textrm{(de,0)}}}}\right)  }\right)  ^{-\frac{\delta}{w+\delta}}.
\end{equation}
We observe that for $\delta=0$ and $w=-1,$ the solution for the $\Lambda$CDM
is recovered and the parameter $\hat{\Omega}_{{\textrm{(dm,0)}}}={\Omega}_{{\textrm{(dm,0)}}}$ is the energy density for the dark matter. In the presence of a baryonic fluid, with energy density $\Omega_{\textrm{b}}\left(
z\right)  =\Omega_{\textrm{(b,0)}}\left(  1+z\right)  ^{3}$, which does not interact with
the rest of the cosmic fluid, the above exact solution for the dark
matter-dark energy is modified. However, because the contribution of the
baryons is small, we can asymptotically describe the Hubble function as \cite{vanderWesthuizen:2025III}:
\begin{equation}
 E_{C}%
^{2}\left(  z\right) \equiv\Omega_{\textrm{(b,0)}}\left(  1+z\right)  ^{3}+\left(  \hat{\Omega}_{{\textrm{(dm,0)}}}\left(  1+z\right)  ^{3\left(  1-\delta\right)  }+\Omega_{{\textrm{(de,0)}}}\left(  1+z\right)  ^{3\left(  1+w\right)  }\right)  \left(
\frac{1+\left(  \frac{\hat{\Omega}_{{\textrm{(dm,0)}}}}{\Omega_{{\textrm{(de,0)}}}}\right)  \left(
1+z\right)  ^{-3\left(  w+\delta\right)  }}{1+\left(  \frac{\hat{\Omega
}_{{\textrm{(dm,0)}}}}{\Omega_{{\textrm{(de,0)}}}}\right)  }\right)  ^{-\frac{\delta}{w+\delta}%
},\label{NLID1_dm_de_BG}%
\end{equation}
with constraint $\frac{H\left(  z\right)  }{H_{0}} \Big|_{z\rightarrow0}=1,$ that
is,
\begin{equation}
\Omega_{{\textrm{(de,0)}}}=1-\hat\Omega_{{\textrm{(dm,0)}}}-\Omega_{\textrm{(b,0)}}\;,%
\end{equation}
which we use in the following Section in order to perform our data analysis and constrain the free parameters.

%\textcolor{blue}{
It has recently been argued that the phantom crossing suggested by the DESI  collaboration \cite{DESI:2025fii}, may be due to an effective dark energy equation of state, that arises due dark sector interactions \cite{guedezounme2025phantomcrossingdarkinteraction}. We investigate this possibility for our model as well. The interaction between the dark sectors leads to both dark matter and dark energy exhibiting an effective equation of state, which can be understood to be an equivalent description of a non-interacting model with $Q=0$ and a dynamical equation of state $w_{\rm{dm}}=w_{\rm{dm}}^{\rm{eff}}$ and $w_{\rm{de}}=w_{\rm{de}}^{\rm{eff}}$ in \eqref{conservationEqdm} and \eqref{conservationEqde}, respectively.  The evolution of the effective equations of state are given by \cite{vanderWesthuizen:2025II}:
\begin{gather} \label{omega_eff_dm_de}
\begin{split}
w^{\rm{eff}}_{\rm{dm}}(z) =- \frac{\delta}{ \left(  \frac{\hat{\Omega}_{{\textrm{(dm,0)}}}}{\Omega_{{\textrm{(de,0)}}}}\right)(1+z)^{-3 \left[ w + \delta\right]}+1}  \quad \text{and} \quad w^{\rm{eff}}_{\rm{de}} (z) &= w + \frac{\delta}{1+\left(  \frac{\hat{\Omega
}_{{\textrm{(dm,0)}}}}{\Omega_{{\textrm{(de,0)}}}}\right)(1+z)^{3 \left[ w + \delta\right]} }.
\end{split}
\end{gather}
A second way to investigate the presence of a phantom crossing without dynamical dark matter is to model the dynamics of the compartmentalization as a reconstructed dynamical dark energy equation of state, as done in \cite{M.B.Gavela_2009, vanderWesthuizen:2025II}. The normalised Hubble parameter $E^2_C(z)$ for a dynamical dark energy model with equation of state $\tilde{w}(z)$ in a flat universe without any interactions in the dark sector is given by: 
\begin{gather} \label{wz_1}
\begin{split}
E^2_C(z)&= \Omega_{\text{(r,0)}}(1+z)^{4}+ \Omega_{\text{(bm,0)}}(1+z)^{3}+ \Omega_{\text{(dm,0)}}(1+z)^{3}+   \Omega_{\text{(de,0)}} \text{exp}\left[ 3 \int_0^z dz' \frac{1+\tilde{w}(z')}{1 + z'} \right], 
\end{split}
\end{gather}
Equating \eqref{wz_1} with \eqref{NLID1_dm_de_BG}, and solving for $\tilde{w}(z)$ after differentiating leads to Eq. \eqref{tilde_w}, also found in 
\cite{vanderWesthuizen:2025II}. 
\begin{gather} \label{tilde_w}
\begin{split}
\tilde{w}(z)&=   \frac{ w  }{ 1+ \left(\frac{\hat\Omega_{\text{(dm,0)}}}{\Omega_{\text{(de,0)}}} \right)  (1+z)^{-3(\delta +w)}   - \left(\frac{\hat\Omega_{\text{(dm,0)}}}{\Omega_{\text{(de,0)}}} \right)  (1+z)^{-3w} \left[\frac{1+\left(\frac{\hat\Omega_{\text{(dm,0)}}}{\Omega_{\text{(de,0)}}} \right)  (1+z)^{-3\left( w + \delta\right) }}{1+\left(\frac{\hat\Omega_{\text{(dm,0)}}}{\Omega_{\text{(de,0)}}} \right) } \right]^{\frac{\delta}{(w +\delta) }}  }.
\end{split}
\end{gather}
We may note that \eqref{wz_1} and \eqref{tilde_w} return to the uncoupled case when $\delta=0$, such that $w^{\rm{eff}}_{\rm{dm}}=0$ and $ w^{\rm{eff}}_{\rm{de}}=\tilde{w}(z)=w$. Additionally, at present $\tilde{w}(0)=w$, and in the asymptotic past $(z\rightarrow\infty)$  we have $w^{\rm{eff}}_{\rm{dm}}=0$, $w^{\rm{eff}}_{\rm{de}}=w+\delta$, $\tilde{w}=  0$, and for the asymptotic future $(z\rightarrow-1)$ we have $w^{\rm{eff}}_{\rm{dm}}=-\delta$, $w^{\rm{eff}}_{\rm{de}}=w$ and $ \tilde{w}=  w$.

\section{Observational Constraints}\label{section3}

Given the theoretical model presented in the previous section, we now focus on the observational aspect of the analysis. Specifically, we aim to identify the range of values permitted for the interaction parameter based on recent cosmological data, and to examine how such an interaction influences the constraints on the other cosmological parameters. To do so, we first outline the methodology employed and describe the observational data sets utilised. Subsequently, we present the resulting constraints within the context of the proposed interacting scenario.

\subsection{Methodology}
For the statistical analysis, we use  the Bayesian Markov chain Monte Carlo approach. In particular, we use the publicly  available software \texttt{Monte Python} \cite{Brinckmann:2018cvx} while the cosmological model has been implemented in  \texttt{CLASS} code \cite{CLASS1,CLASS2} by modifying the energy density background equations of dark matter and dark energy with the expressions displayed at equations~\eqref{eq01} to analyze this particular model, along with the standard scenario represented by the $\Lambda$CDM model. 

In our analysis, we consider as free parameters the present-day interacting dark matter density as $\hat{\Omega}_{\textrm{(dm,0)}  }$\footnote{The $\;\hat{ }\;$ denotes the interacting case to reinforce the difference between dark matter and interacting dark matter. In the non-interacting case it would just represent the usual dark matter density of today}, the Hubble constant $H_{0}$,  the dark energy equations of state $w$ and the interacting parameter $\delta$. We include the nuisance parameter $M$ when using supernovae data. As a derived parameter, we consider the present-day total matter density $\Omega_{\textrm{(m,0)}  }=\hat{\Omega}_{\textrm{(dm,0)}  } +\Omega_{\textrm{(b,0)}}$.  We fix our baryonic cosmology to satisfy $\Omega_{\textrm{(b,0)}}=0.0468$, while the neutrino sector is considered to be formed by three massive neutrinos satisfying $\sum m_{\nu}=0.06eV$ and $N_{eff}=3.044$. 
For the previous parameters, we consider flat priors satisfying in each case  $\hat{\Omega}_{\textrm{(dm,0)}  }\in\lbrack0.001,0.9]$, $H_{0}\in\lbrack40,100]$, $w\in
\lbrack-2.00,-0.33]$ and   $M\in\lbrack-30,-10]$. For the model parameter $\delta$, and provided we have no previous knowledge of the allowed values, we consider a flat unbound prior as $\delta\in\lbrack-\infty,+\infty]$.

\subsection{Data}
In this study,  the following two recent datasets are considered  to constrain the model's parameters. The choice of only using late-time probes come down to the fact that the interaction kernel in \eqref{in.01} causes the greatest changes in the dynamics of the universe at low redshifts, but more importantly, including early-time data sets such as CMB require a full perturbation implementation which presently falls outside of the scope of this study.  
\begin{itemize}
\item Pantheon+ Supernova: This dataset includes 1701 light curves of 1550
spectroscopically confirmed supernova events within the range of redshift $0.001%
<z<2.27~$\cite{pan}. We consider the Pantheon+ data with the Supernova $H_{0}$
for the Equation of State of dark energy without the Cepheid calibration (SN). The Pantheon+ dataset is currently the most complete and widely used SN compilation. The luminosity distance~$D_{L}=c\left(  1+z\right)  \int\frac{dz}{H\left(
z\right)  }$, is used to define the theoretical distance modulus$~\mu
^{th}=5\log D_{L}+25$, which is used to constrain with the distance modulus
$\mu^{obs}~$at~observed redshifts~$z$.

\item Baryonic acoustic oscillations (BAO) distance 
of the 
recent data from the DESI DR2 release \cite{DESI:2025zgx,DESI:2025fii,DESI:2025qqy,DESI:2025wyn}. The dataset
includes observation values of the $\frac{D_{M}}{r_{d}}=\frac{\left(
1+z\right)  ^{-1}D_{L}}{r_{d}},~\frac{D_{V}}{r_{d}}=\frac{\left(  cD_{L}%
\frac{z}{H\left(  z\right)  }\right)  ^{\frac{1}{3}}}{r_{d}}$~and$~\frac
{D_{H}}{r_{d}}=\frac{c}{r_{d}H}$ where $r_{d}$ is the sound horizon at the
drag epoch, $D_{M}$ is the comoving angular distance; $D_{V}$ is the volume
averaged distance and $D_{H}$ is the Hubble distance.
\end{itemize}

\subsection{Results}
\begin{figure}[h]
\centering\includegraphics[width=1\linewidth]{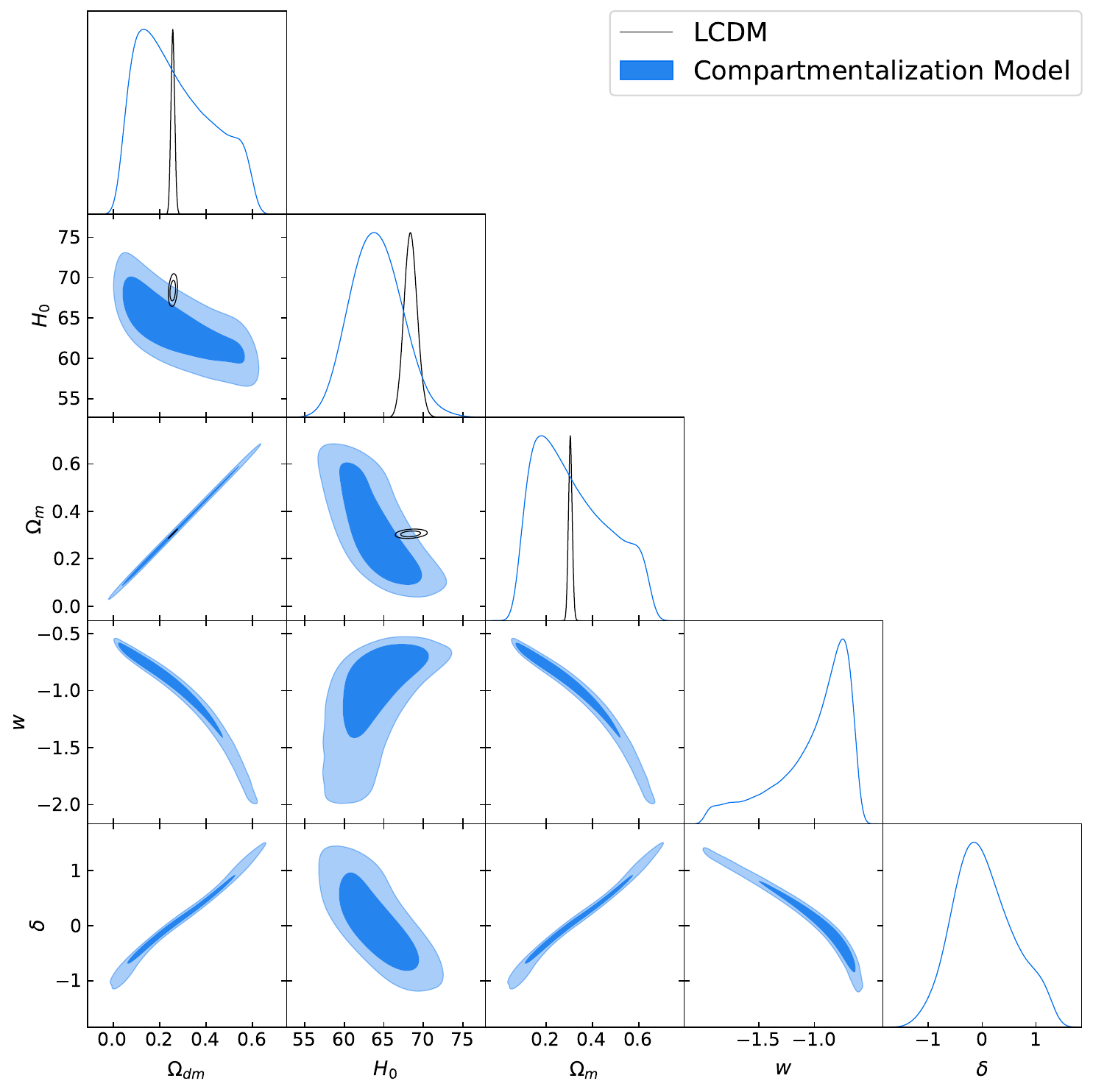}
\label{Fig:MP_non_linear_3-2}\caption{Contour plots for  Pantheon+ SNIa \& DESI DR2 BAO  data.}%
\label{fig1}%
\end{figure}

In Table~\ref{tab1} and \ref{tab2}, we display the constraints obtained for the $\Lambda$CDM, the $w_0w_a$CDM, and the Compartmentalization model on the cosmological, nuisance and derived parameters, using BAO data alone and combined BAO and SNIa data, respectively. In Figure~\ref{fig1}, we also display the corresponding one-dimensional and two-dimensional contours for both cases. From the results for the interaction model from both datasets on the coupling parameter  $\delta= -0.23\pm{0.76}$ and  $\delta= 0.04_{-0.66}^{+0.47}$, we can observe the non-interacting case, $\delta=0$, is within the  $1\sigma$ confidence level. We also obtain a value of the equation of state of dark energy compatible with a non-dynamical dark energy $w=-1$ in the $1\sigma$ confidence level. The other cosmological parameters tend to have larger uncertainties than in the $\Lambda$CDM and $w_0w_a$CDM, which from Figure~\ref{fig1} can be understood due to the strong degeneracy between the coupling parameter $\delta$ and $H_0$ or $\Omega_{\rm{dm}}$.  

\begin{figure}[htbp]
    \centering
    \begin{subfigure}[b]{0.49\linewidth}
        \centering
        \includegraphics[width=\linewidth]{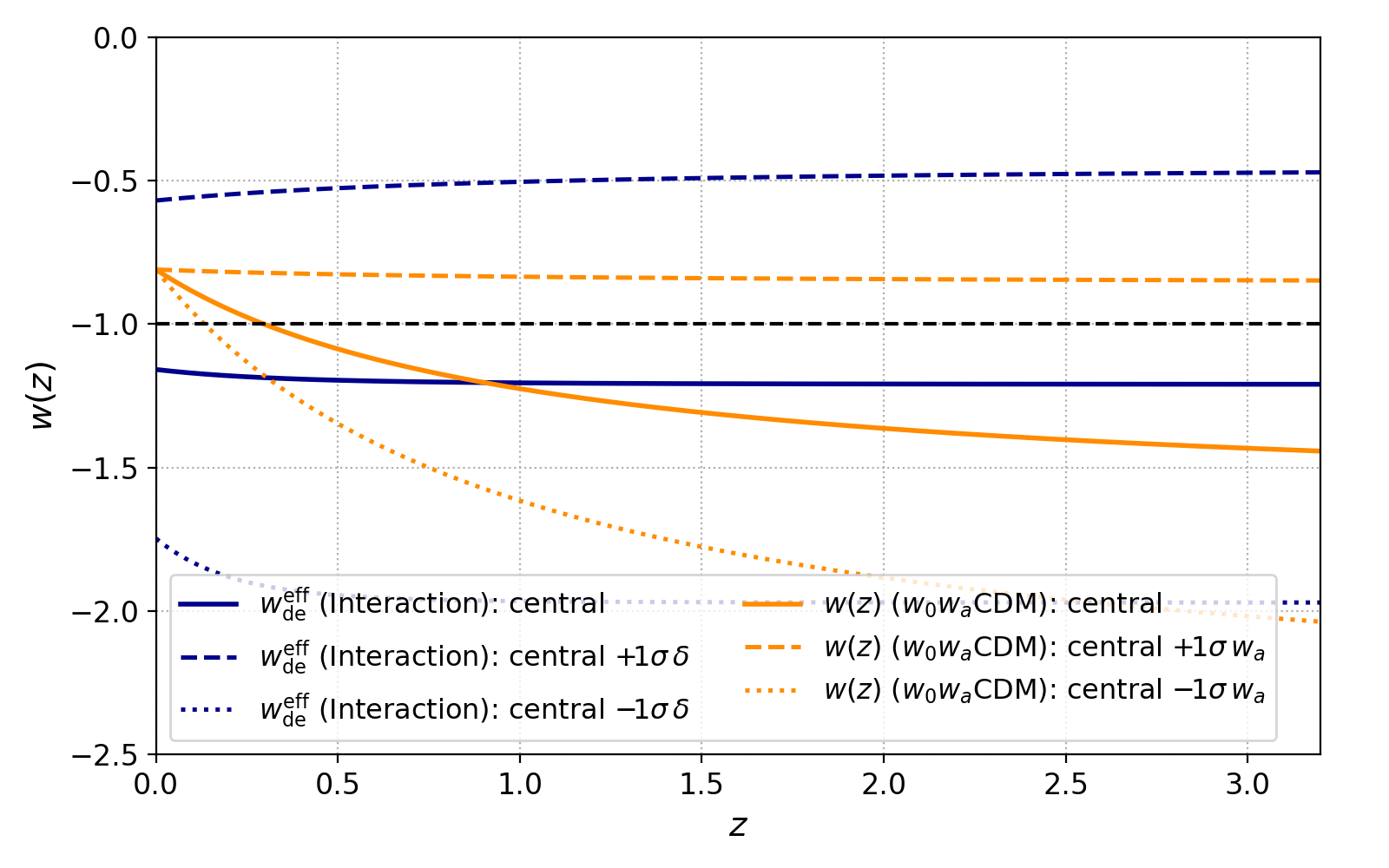}
        \label{fig:eos_BAO}
    \end{subfigure}%
    \hspace{0pt} % No extra space between subfigures
    \begin{subfigure}[b]{0.49\linewidth}
        \centering
        \includegraphics[width=\linewidth]{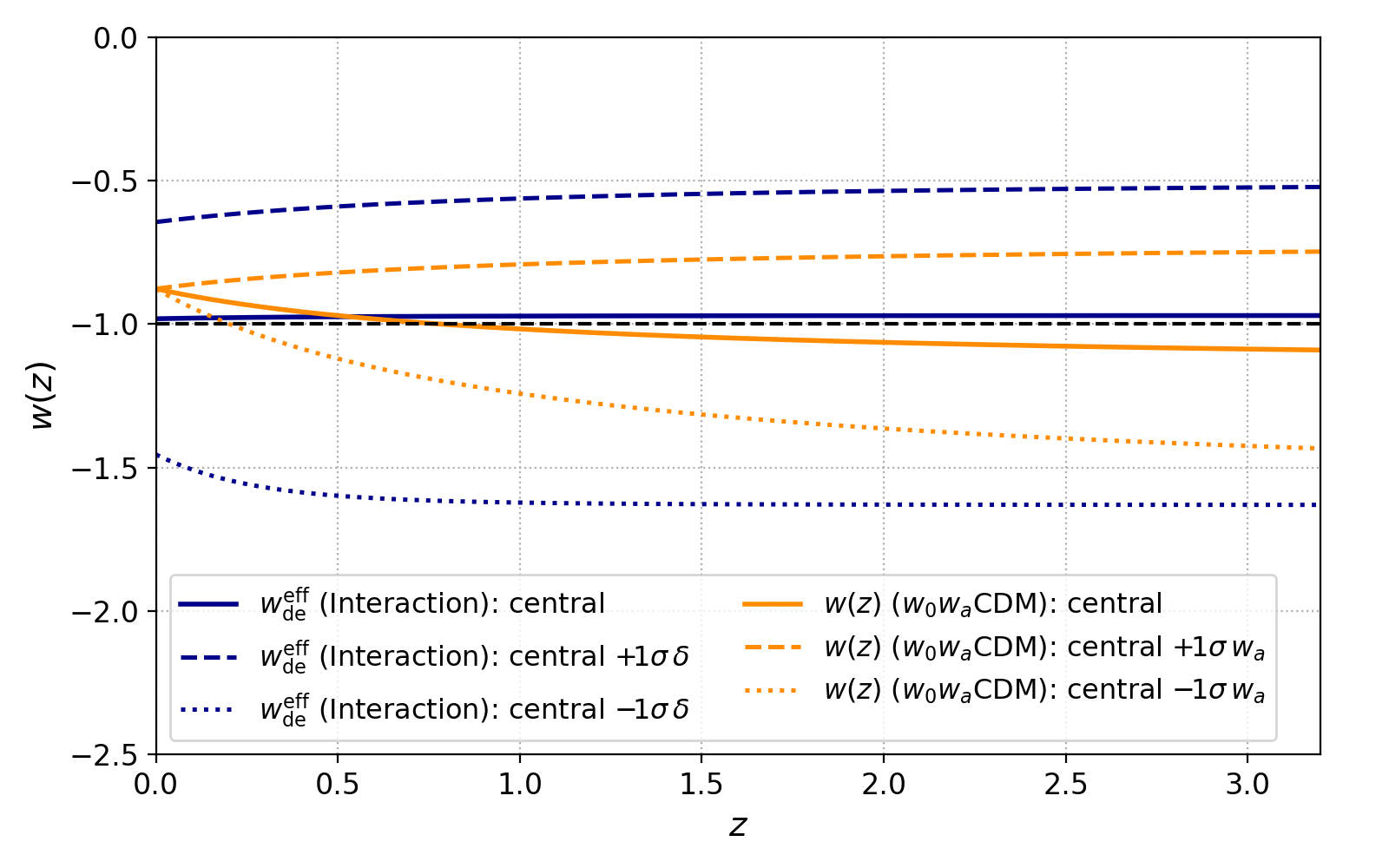}
        \label{fig:eos_BAOSN1}
    \end{subfigure}    
    \caption{Evolution of $w(z)$ for the $w_0 w_a$CDM model, and the effective equation of state $w^{\rm{eff}}_{\rm{de}}(z)$ for the interacting dark energy model. For clarity of illustration, central values are plotted alongside the $\pm1 \sigma$ variation in only $w_a$ and $\delta$, showcasing how the uncertainties in the posteriors can affect $w(z)$ and $w^{\rm{eff}}_{\rm{de}}(z)$. The left panel is produced using only BAO data from Table \ref{tab1}, while the right panel includes BAO+SNIa data from Table \ref{tab2}.}
    \label{fig:w_eff}
\end{figure}

\begin{figure}[htbp]
    \centering
    \begin{subfigure}[t]{0.49\linewidth}
        \centering
        \includegraphics[width=\linewidth]{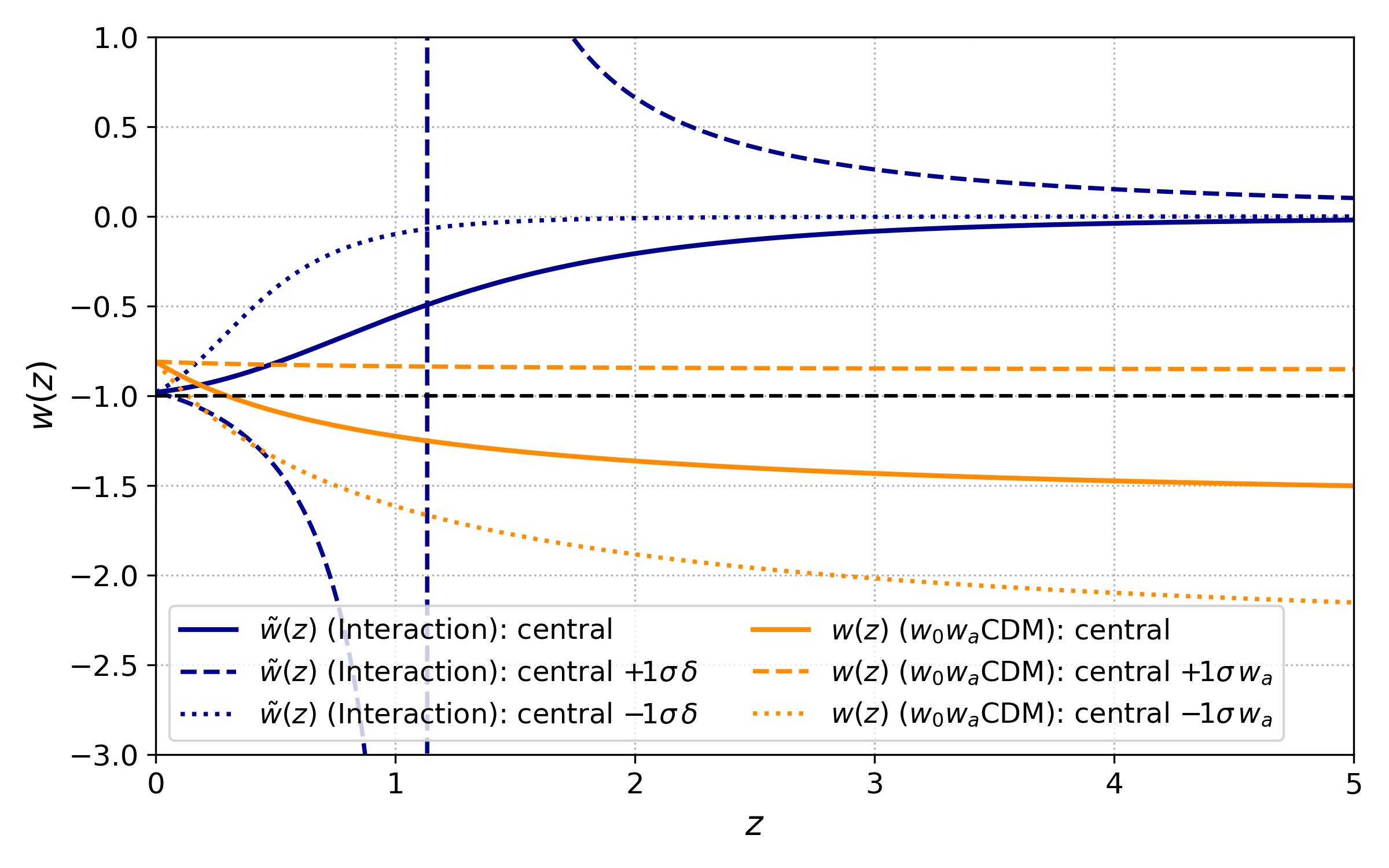}
        \label{fig:eos_BAO}
    \end{subfigure}%
    \hspace{0pt} % No extra space between subfigures
    \begin{subfigure}[t]{0.49\linewidth}
        \centering
        \includegraphics[width=\linewidth]{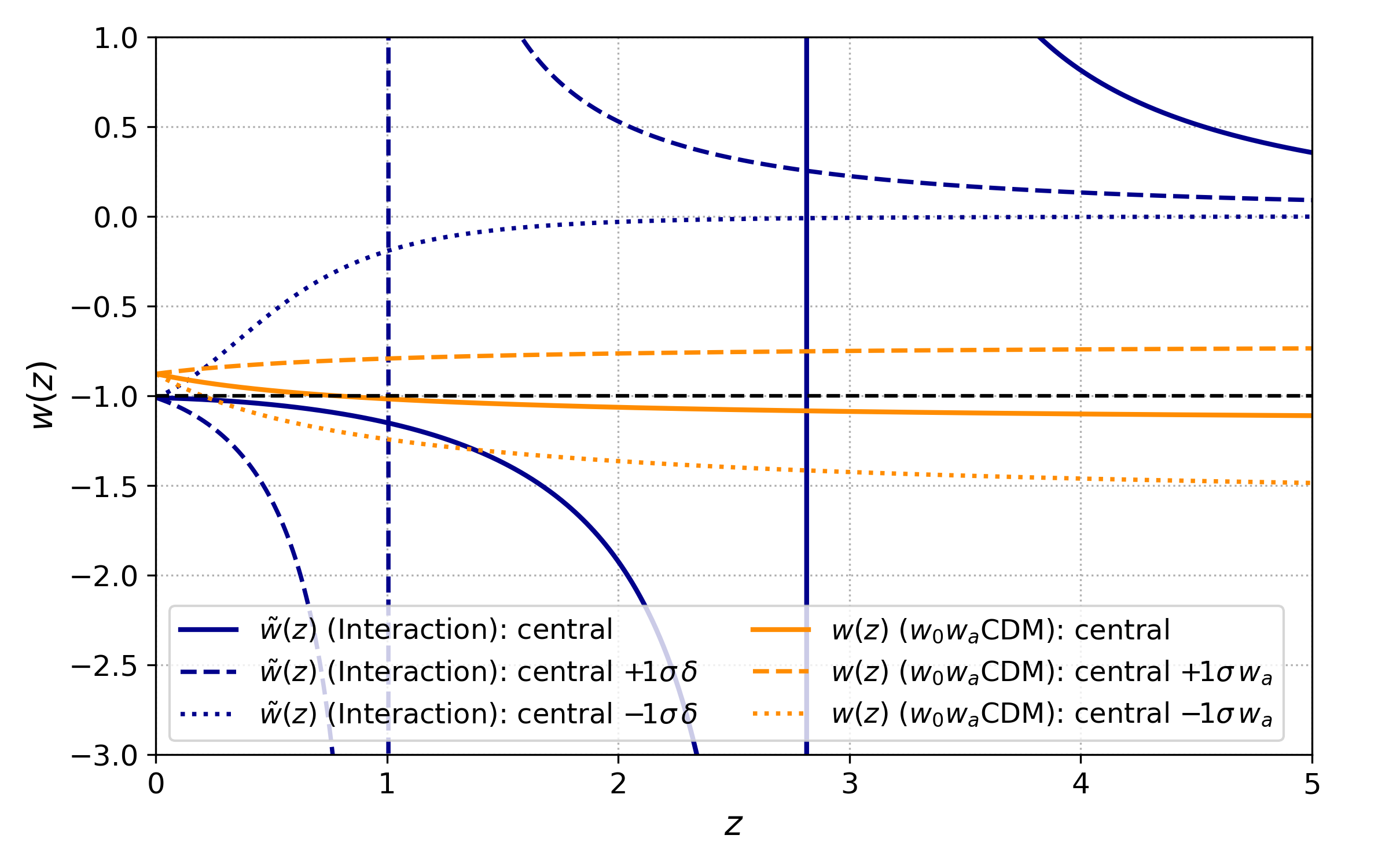}
    \end{subfigure}    
    \caption{Evolution of $w(z)$ for the $w_0 w_a$CDM model, and reconstructed equation of state $\tilde{w}(z)$ for the interacting dark energy model. Central values are plotted alongside the $\pm1 \sigma$ variation in only $w_a$ and $\delta$. Divergent behaviour is shown when energy flows from dark energy to dark matter ($\delta>0$), as discussed for this specific interaction in Appendix C and Figure 16 of \cite{vanderWesthuizen:2025II}. The left panel is produced using only BAO data from Table \ref{tab1}, while the right panel includes BAO+SNIa data from Table \ref{tab2}}
    \label{fig:tilde_w}
\end{figure}

For the $w_0w_a$CDM, from both datasets we obtained $w_0>-1$ within the  $1\sigma$ confidence level, indicating dark energy that is within the quintessence regime and weakening at present, which is in agreement with \cite{DESI:2025fii}. The possibility of a dark energy phantom crossing for the $w_0w_a$CDM and the interacting model is illustrated by plotting the evolution of $w(z)=w_0+w_a(1-a)$ with $a=(1+z)^{-1}$, and both $w_{\rm{de}}^{\rm{eff}}(z)$ and $\tilde{w}(z)$ from \eqref{omega_eff_dm_de} and \eqref{tilde_w}, respectively. This is done using the posteriors obtained from BAO data (Table \ref{tab1}) and BAO+SNIa (Table \ref{tab2}) and is plotted in Figure \ref{fig:w_eff} and \ref{fig:tilde_w}. Here we see that a phantom crossing is allowed for both models, given the $\pm1\sigma$ interval of $w_a$ and $\delta$ from the central values. We have chosen to vary only one parameter for clarity of illustration, but readers should be aware that more complex plots can be obtained by varying the other parameters in Table \ref{tab1} and \ref{tab2}. For the interacting model, a cosmological constant description is not excluded as the uncertainties are too large to draw any insightful conclusions.

From the combined BAO and SNIa data set in Table \ref{tab2} we found for both the interacting case and $w_0w_a$CDM that $ \chi_{w_0 w_a}^{2}=\chi_{\text{Interaction}}^{2}=1418$, while for the $\Lambda$CDM model we obtain $\chi_{\Lambda\text{CDM}}^{2}=1423$, which means that the comparison between the scenarios gives $\chi_{w_0 w_a /\text{Interaction}}^{2}-\chi_{\Lambda\text{CDM}}^{2}=-5$. For BAO alone in Table \ref{tab1}, this preference is also shown as $\chi_{w_0 w_a}^{2}-\chi_{\Lambda\text{CDM}}^{2}=-1.3$ and $\chi_{\text{Interaction}}^{2}-\chi_{\Lambda\text{CDM}}^{2}=-2.8$.

Provided both scenarios have different degrees of freedom, in order to compare them, we use the Akaike Information Criterion~$\text{AIC}=\chi^{2}+2~\text{d.o.f.}$~\cite{AIC}.  We calculate the difference to be $\text{AIC}_{w_0 w_a /\text{Interaction}}-\text{AIC}_{\Lambda\text{CDM}}=-1$, which means that there is weak evidence in favor of both the $w_0 w_a$CDM model and the interacting case with respect the concordance model. The preference for both cases is reduced when considering BAO alone, as we obtain $\text{AIC}_{w_0 w_a} -\text{AIC}_{\Lambda\text{CDM}}=2.7$ and $\text{AIC}_{\text{Interaction}}-\text{AIC}_{\Lambda\text{CDM}}=1.2$, thus indicating a preference for the $\Lambda$CDM model. Lastly, comparing our interaction model with $w_0 w_a$CDM, which is the natural two parameter extension to $\Lambda$CDM, we find no difference in the fitting for the combined BAO and SNIa, $\text{AIC}_{\text{Interaction}}-\text{AIC}_{w_0 w_a} =0$, but a preference for the interaction model when considering BAO alone, $\text{AIC}_{\text{Interaction}}-\text{AIC}_{w_0 w_a} =-1.5$.
In addition, we have to bear in mind that the coupling parameter $\delta$ is poorly constrained and, even more, the non-interacting case is included in the $1\sigma$ confidence level.

\begin{table}[tbp] \centering
%EndExpansion
\begin{tabular}[c]{cccccc}
\hline%\hline  
\multicolumn{6}{c}{BAO alone}\\
\multicolumn{2}{c}{ \textbf{Parameter} $\Lambda$\textbf{CDM} } & \multicolumn{2}{c}{ \textbf{Parameter} $w_0 w_a$\textbf{CDM} }& \multicolumn{2}{c}{
\textbf{Parameter} \textbf{Interaction}}\\\hline
\centering
{\boldmath $\Omega{}_{{\textrm{\textbf{(dm,0)}}}}$} & $0.2485\pm 0.0083$ & {\boldmath $\Omega{}_{{\textrm{\textbf{(dm,0)}}}}$} & $0.274^{+0.028}_{-0.019}$ &
{\boldmath$\hat{\Omega}{}_{{\textbf{\textrm{(dm,0)}}}}$} & $ 0.215^{+0.093}_{-0.215}  $\\
{\boldmath$\Omega{}_{\left(  m,0\right)  }$} & $ 0.2967\pm 0.0083$& {\boldmath$\Omega{}_{\left(  m,0\right)  }$}  &  $0.322^{+0.028}_{-0.019}$ &
{\boldmath$\Omega{}_{\left(  m,0\right)  }$} & $ 0.263^{+0.093}_{-0.22}  $\\
&  &{\boldmath $w_0$ }& $ -0.81^{+0.18}_{-0.16}$ &   {\boldmath$w$} & $ -0.98^{+0.47}_{-0.16}     $\\
&  &{\boldmath $w_a$ }&  $-0.83\pm 0.78$ & {\boldmath$\delta{}$} & $  -0.23\pm 0.76   $\\\hline%\hline
{\boldmath$\chi_{\Lambda\text{\textbf{CDM}}}^{2}$ }& $10.5$ & {\boldmath$\chi_{w_0 w_a}^{2}$} & $ 9.2 $\ & {\boldmath$\chi_{\text{\textbf{Interaction}}}^{2}$ }& $7.7$\\%\hline
{\boldmath$\rm{AIC}_{\Lambda\text{\textbf{CDM}}}$ }& $14.5$ & {\boldmath$\text{AIC}_{w_0 w_a}$ } & $ 17.2 $\ & {\boldmath$\text{AIC}_{\rm{Interaction}}$ } & $15.7$\\\hline%\hline
\end{tabular}
\caption{Constraints on the cosmological parameters using DESI DR2 BAO data.}%
\label{tab1}%
%TCIMACRO{\TeXButton{E}{\end{table}}}%
%BeginExpansion
\end{table}%

%TCIMACRO{\TeXButton{B}{\begin{table}[tbp] \centering}}%
%BeginExpansion
\begin{table}[tbp] \centering
%EndExpansion
\begin{tabular}[c]{cccccc}
\hline%\hline  
\multicolumn{6}{c}{BAO \& SNIa}\\
\multicolumn{2}{c}{ \textbf{Parameter} $\Lambda$\textbf{CDM} } & \multicolumn{2}{c}{ \textbf{Parameter} $w_0 w_a$\textbf{CDM} }& \multicolumn{2}{c}{
\textbf{Parameter} \textbf{Interaction}}\\\hline
\centering
{\boldmath $\Omega{}_{{\textrm{\textbf{(dm,0)}}}}$} & $0.2569\pm0.0080$ & {\boldmath $\Omega{}_{{\textrm{\textbf{(dm,0)}}}}$} &  $0.256^{+0.022}_{-0.011}$  &
{\boldmath$\hat{\Omega}{}_{{\textbf{\textrm{(dm,0)}}}}$} & $0.270_{-0.22}^{+0.098}%
$\\
{\boldmath$H_{0}$} & $68.42_{-0.87}^{+0.78}$ &  {\boldmath$H_{0}$} & $ 65.1^{+3.8}_{-2.8} $  & {\boldmath$H_{0}$} &
$64.1_{-3.5}^{+3.0}$\\
{\boldmath$M$} & $-19.410\pm0.028$ & {\boldmath$M$} & $-19.51^{+0.14}_{-0.087}$   & {\boldmath$M$} & $-19.54\pm0.11$\\
{\boldmath$\Omega{}_{\left(  m,0\right)  }$} & $0.3051\pm0.0080$ &  {\boldmath$\Omega{}_{\left(  m,0\right)  }$} & $0.304^{+0.021}_{-0.010}$ &
{\boldmath$\Omega{}_{\left(  m,0\right)  }$} & $0.318_{-0.22}^{+0.098}$\\
&  & {\boldmath $w_0$ }& $-0.877^{+0.056}_{-0.065}$ &  {\boldmath$w$} & $-1.01_{-0.11}^{+0.39}$\\
&  & {\boldmath $w_a$ }& $-0.28\pm0.45$ &  {\boldmath$\delta{}$} & $0.04_{-0.66}^{+0.47}$\\\hline%\hline
{\boldmath$\chi_{\Lambda\text{\textbf{CDM}}}^{2}$ }&$1423$ & {\boldmath$\chi_{w_0 w_a}^{2}$}&$ 1418$ & {\boldmath$\chi_{\text{\textbf{Interaction}}}^{2}$ }& 1418\\%
{\boldmath$\rm{AIC}_{\Lambda\text{\textbf{CDM}}}$ }& $1429$ & {\boldmath$\text{AIC}_{w_0 w_a}$ } & $ 1428 $\ & {\boldmath$\text{AIC}_{\rm{Interaction}}$ } & $1428$\\\hline%\hline
\end{tabular}
\caption{Constraints on the cosmological parameters for $\Lambda$CDM, $w_0 w_a$CDM and an interacting model using Pantheon+ SNIa \& DESI DR2 BAO data.}%
\label{tab2}%
%TCIMACRO{\TeXButton{E}{\end{table}}}%
%BeginExpansion
\end{table}%
%EndExpansion

\section{Conclusions}\label{section4}
In this analysis, we examined a non-linear interacting model using recent cosmological data relevant to the late-time universe. Specifically, we considered an interaction between the dark energy and dark matter sectors, leading to a system of differential equations that, analogously to ecological models, describes a form of compartmentalization between the two "species". A key feature of this gravitational model is that it admits an analytic expression for the Hubble parameter.

As far as the statistical analysis is concerned, in this work we use recent BAO data from DESI DR2~\cite{DESI:2025zgx,DESI:2025fii,DESI:2025qqy,DESI:2025wyn} and the Pantheon+ compilation~\cite{pan}, excluding the SH0ES calibration, to constrain both the standard $\Lambda$CDM scenario and the interacting (compartmentalized) model. From the resulting constraints on the interaction parameter, $\delta = 0.04^{+0.47}_{-0.66}$, we find that the interacting model cannot be statistically distinguished from the non-interacting case. In particular, our results reveal a strong correlation between $\delta$ and the other cosmological parameters. In particular, when energy flows from dark energy to dark matter ($\delta > 0$), the data favor higher values of the Hubble constant $H_0$ and an increased dark matter density $\Omega_{\rm dm}$. Conversely, when energy is transferred from dark matter to dark energy ($\delta < 0$), lower values of $H_0$ and $\Omega_{\rm dm}$ are preferred.

Regarding the well-known cosmological tensions, in particular here for the $H_0$ crisis~\cite{Abdalla:2022yfr}, the proposed interaction has the potential to shift the inferred value of the Hubble constant towards one that is more compatible with Planck CMB measurements~\cite{Planck:2018vyg}. However, based on the current results presented in Table~\ref{tab1}, we find that the value of $H_0$ exhibits larger uncertainties compared to the standard $\Lambda$CDM case. As a result, it remains inconclusive whether the tension is significantly alleviated in this scenario.  
%\textcolor{blue}{
The possibility of the dark energy effective equation of state $w_{\rm{de}}^{\rm{eff}}$ and dynamical dark energy description $\tilde{w}(z)$ having a phantom crossing is permitted within uncertainties, but not required. The uncertainties on our posteriors are too large to make any definitive claims, as seen in Figure \ref{fig:w_eff} and \ref{fig:tilde_w}. In Figure \ref{fig:tilde_w}, we can also see divergent behaviour occurring for $\tilde{w}(z)$, which is expected if energy flows from dark energy to dark matter, as previously noted in \cite{M.B.Gavela_2009}, and further discussed for this specific interaction in Appendix C and Figure 16 of \cite{vanderWesthuizen:2025II}. There it is shown that this divergence is a feature of the parameterization and not a pathology in the underlying dynamics of the model itself.

Furthermore, the interaction may be free from large-scale instabilities, as the doom factor requirement that $(1+w)$ and $\delta$ have opposite signs is met for the best fit values for both datasets. Unfortunately, this is not guaranteed within the 68 \% confidence interval. In order to address this more confidently, a full perturbation study for the interaction will be done in a future work. For now we refer the reader to previous perturbation studies found in \cite{M.B.Gavela_2009, Honorez_2010, Li_2014}.

It is worth noting that our constraints prefer a phantom-regime dark energy equation of state, $w=-1.01_{-0.11}^{+0.39}$, which if taken at face value for this interaction would predict a big rip singularity in the distant future \cite{vanderWesthuizen:2025II}.  Nevertheless, this result is not statistically significant, as a dark energy equation of state in the quintessence regime (which avoids a big rip) is also within the $1\sigma$ confidence level. This is purely a phenomenological observation, and the model remains fully compatible with late-time observational data.

We also examined the behaviour of the interacting model when SH0ES data are included. We have found there is a preference for non-vanishing values of the interacting parameter when the SH0ES calibration is used for the Pantheon+ dataset as have been already seen in previous works~\cite{Pan:2025qwy}. In particular, a larger value of the Hubble constant $H_0$ is preferred, which necessarily selects negative values of the interacting parameter as we can even infer from Figure~\ref{fig1}. The effect of SH0ES calibration is, then, the consequence of the strong degeneracy between the interacting parameter and the other cosmological parameters and, because of that, we present here only the results without using the SH0ES calibration. 
It may be mentioned that for supernovae data, we used only Pantheon+, as it is the most complete supernova data set, but preliminary checks with both DES 5YR and Union3 found consistent posteriors within uncertainties.

Last but not least, regarding the statistical preference of the compartmentalization model with respect to both the $\Lambda$CDM and $w_0w_a$CDM, although the fit improves slightly, we find only weak preference for the model when using the AIC criteria, provided $\text{AIC}_{\text{Interaction}}-\text{AIC}_{\Lambda\text{CDM}}=-1$ and $\text{AIC}_{\text{Interaction}}-\text{AIC}_{w_0 w_a} =0$ when considering BAO and SNIa together. This preference depends on the dataset and is not present when considering BAO alone, which yields $\text{AIC}_{\text{Interaction}}-\text{AIC}_{\Lambda\text{CDM}}=1.2$, but this dataset still presents a preference for the interaction over $w_0w_a$CDM, as $\text{AIC}_{\text{Interaction}}-\text{AIC}_{w_0 w_a} =-1.5$. 

We conclude that non-linear interacting models can yield cosmological scenarios that are consistent with observational data. Although this work focused on a phenomenological interaction, it is worthwhile to investigate the origin of such interactions within a theoretical framework. The interaction kernel studied here most strongly deviates from the concordance model at low redshift, for which we have focused on late-time probes. Further investigations that include more datasets, especially early-time datasets such as CMB, will be crucial for further works, but require a full perturbation implementation, which will be considered in a future work. Once this is done, we may obtain more stringent constraints on the possibility of compartmentalization in the dark sector of the universe.  

\begin{acknowledgments}
 DF acknowledged support from "Convocatoria de contratación de personal investigador doctor de la UPV/EHU (2024)".
 AP thanks the support of Vicerrector\'{\i}a de Investigaci\'{o}n y Desarrollo. Tecnol\'{o}gico (Vridt) at Universidad Cat\'{o}lica del Norte through N\'{u}cleo de Investigaci\'{o}n Geometr\'{\i}a Diferencial y Aplicaciones, Resoluci\'{o}n Vridt No - 096/2022. AP was partially supported by Proyecto
Fondecyt Regular 2024, Folio 1240514, Etapa 2025. 
\end{acknowledgments}

\bibliography{biblio}  

%apsrev4-2.bst 2019-01-14 (MD) hand-edited version of apsrev4-1.bst
%Control: key (0)
%Control: author (8) initials jnrlst
%Control: editor formatted (1) identically to author
%Control: production of article title (0) allowed
%Control: page (0) single
%Control: year (1) truncated
%Control: production of eprint (0) enabled
\begin{thebibliography}{70}%
\makeatletter
\providecommand \@ifxundefined [1]{%
 \@ifx{#1\undefined}
}%
\providecommand \@ifnum [1]{%
 \ifnum #1\expandafter \@firstoftwo
 \else \expandafter \@secondoftwo
 \fi
}%
\providecommand \@ifx [1]{%
 \ifx #1\expandafter \@firstoftwo
 \else \expandafter \@secondoftwo
 \fi
}%
\providecommand \natexlab [1]{#1}%
\providecommand \enquote  [1]{``#1''}%
\providecommand \bibnamefont  [1]{#1}%
\providecommand \bibfnamefont [1]{#1}%
\providecommand \citenamefont [1]{#1}%
\providecommand \href@noop [0]{\@secondoftwo}%
\providecommand \href [0]{\begingroup \@sanitize@url \@href}%
\providecommand \@href[1]{\@@startlink{#1}\@@href}%
\providecommand \@@href[1]{\endgroup#1\@@endlink}%
\providecommand \@sanitize@url [0]{\catcode `\\12\catcode `\$12\catcode
  `\&12\catcode `\#12\catcode `\^12\catcode `\_12\catcode `\%12\relax}%
\providecommand \@@startlink[1]{}%
\providecommand \@@endlink[0]{}%
\providecommand \url  [0]{\begingroup\@sanitize@url \@url }%
\providecommand \@url [1]{\endgroup\@href {#1}{\urlprefix }}%
\providecommand \urlprefix  [0]{URL }%
\providecommand \Eprint [0]{\href }%
\providecommand \doibase [0]{https://doi.org/}%
\providecommand \selectlanguage [0]{\@gobble}%
\providecommand \bibinfo  [0]{\@secondoftwo}%
\providecommand \bibfield  [0]{\@secondoftwo}%
\providecommand \translation [1]{[#1]}%
\providecommand \BibitemOpen [0]{}%
\providecommand \bibitemStop [0]{}%
\providecommand \bibitemNoStop [0]{.\EOS\space}%
\providecommand \EOS [0]{\spacefactor3000\relax}%
\providecommand \BibitemShut  [1]{\csname bibitem#1\endcsname}%
\let\auto@bib@innerbib\@empty
%</preamble>
\bibitem [{\citenamefont {Abdul~Karim}\ \emph {et~al.}(2025)\citenamefont
  {Abdul~Karim} \emph {et~al.}}]{DESI:2025zgx}%
  \BibitemOpen
  \bibfield  {author} {\bibinfo {author} {\bibfnamefont {M.}~\bibnamefont
  {Abdul~Karim}} \emph {et~al.} (\bibinfo {collaboration} {DESI}),\ }\bibfield
  {title} {\bibinfo {title} {{DESI DR2 Results II: Measurements of Baryon
  Acoustic Oscillations and Cosmological Constraints}},\ }\href@noop {} {\
  (\bibinfo {year} {2025})},\ \Eprint {https://arxiv.org/abs/2503.14738}
  {arXiv:2503.14738 [astro-ph.CO]} \BibitemShut {NoStop}%
\bibitem [{\citenamefont {Lodha}\ \emph {et~al.}(2025)\citenamefont {Lodha}
  \emph {et~al.}}]{DESI:2025fii}%
  \BibitemOpen
  \bibfield  {author} {\bibinfo {author} {\bibfnamefont {K.}~\bibnamefont
  {Lodha}} \emph {et~al.} (\bibinfo {collaboration} {DESI}),\ }\bibfield
  {title} {\bibinfo {title} {{Extended Dark Energy analysis using DESI DR2 BAO
  measurements}},\ }\href@noop {} {\  (\bibinfo {year} {2025})},\ \Eprint
  {https://arxiv.org/abs/2503.14743} {arXiv:2503.14743 [astro-ph.CO]}
  \BibitemShut {NoStop}%
\bibitem [{\citenamefont {Andrade}\ \emph {et~al.}(2025)\citenamefont {Andrade}
  \emph {et~al.}}]{DESI:2025qqy}%
  \BibitemOpen
  \bibfield  {author} {\bibinfo {author} {\bibfnamefont {U.}~\bibnamefont
  {Andrade}} \emph {et~al.} (\bibinfo {collaboration} {DESI}),\ }\bibfield
  {title} {\bibinfo {title} {{Validation of the DESI DR2 Measurements of Baryon
  Acoustic Oscillations from Galaxies and Quasars}},\ }\href@noop {} {\
  (\bibinfo {year} {2025})},\ \Eprint {https://arxiv.org/abs/2503.14742}
  {arXiv:2503.14742 [astro-ph.CO]} \BibitemShut {NoStop}%
\bibitem [{\citenamefont {Gu}\ \emph {et~al.}(2025)\citenamefont {Gu} \emph
  {et~al.}}]{DESI:2025wyn}%
  \BibitemOpen
  \bibfield  {author} {\bibinfo {author} {\bibfnamefont {G.}~\bibnamefont {Gu}}
  \emph {et~al.} (\bibinfo {collaboration} {DESI}),\ }\bibfield  {title}
  {\bibinfo {title} {{Dynamical Dark Energy in light of the DESI DR2 Baryonic
  Acoustic Oscillations Measurements}},\ }\href@noop {} {\  (\bibinfo {year}
  {2025})},\ \Eprint {https://arxiv.org/abs/2504.06118} {arXiv:2504.06118
  [astro-ph.CO]} \BibitemShut {NoStop}%
\bibitem [{\citenamefont {Adame}\ \emph
  {et~al.}(2025{\natexlab{a}})\citenamefont {Adame} \emph
  {et~al.}}]{DESI:2024uvr}%
  \BibitemOpen
  \bibfield  {author} {\bibinfo {author} {\bibfnamefont {A.~G.}\ \bibnamefont
  {Adame}} \emph {et~al.} (\bibinfo {collaboration} {DESI}),\ }\bibfield
  {title} {\bibinfo {title} {{DESI 2024 III: baryon acoustic oscillations from
  galaxies and quasars}},\ }\href
  {https://doi.org/10.1088/1475-7516/2025/04/012} {\bibfield  {journal}
  {\bibinfo  {journal} {JCAP}\ }\textbf {\bibinfo {volume} {04}},\ \bibinfo
  {pages} {012}},\ \Eprint {https://arxiv.org/abs/2404.03000} {arXiv:2404.03000
  [astro-ph.CO]} \BibitemShut {NoStop}%
\bibitem [{\citenamefont {Adame}\ \emph
  {et~al.}(2025{\natexlab{b}})\citenamefont {Adame} \emph
  {et~al.}}]{DESI:2024mwx}%
  \BibitemOpen
  \bibfield  {author} {\bibinfo {author} {\bibfnamefont {A.~G.}\ \bibnamefont
  {Adame}} \emph {et~al.} (\bibinfo {collaboration} {DESI}),\ }\bibfield
  {title} {\bibinfo {title} {{DESI 2024 VI: cosmological constraints from the
  measurements of baryon acoustic oscillations}},\ }\href
  {https://doi.org/10.1088/1475-7516/2025/02/021} {\bibfield  {journal}
  {\bibinfo  {journal} {JCAP}\ }\textbf {\bibinfo {volume} {02}},\ \bibinfo
  {pages} {021}},\ \Eprint {https://arxiv.org/abs/2404.03002} {arXiv:2404.03002
  [astro-ph.CO]} \BibitemShut {NoStop}%
\bibitem [{\citenamefont {Park}\ \emph
  {et~al.}(2024{\natexlab{a}})\citenamefont {Park}, \citenamefont
  {de~Cruz~P\'erez},\ and\ \citenamefont {Ratra}}]{Park:2024vrw}%
  \BibitemOpen
  \bibfield  {author} {\bibinfo {author} {\bibfnamefont {C.-G.}\ \bibnamefont
  {Park}}, \bibinfo {author} {\bibfnamefont {J.}~\bibnamefont
  {de~Cruz~P\'erez}},\ and\ \bibinfo {author} {\bibfnamefont {B.}~\bibnamefont
  {Ratra}},\ }\bibfield  {title} {\bibinfo {title} {{Using non-DESI data to
  confirm and strengthen the DESI 2024 spatially flat w0waCDM cosmological
  parametrization result}},\ }\href
  {https://doi.org/10.1103/PhysRevD.110.123533} {\bibfield  {journal} {\bibinfo
   {journal} {Phys. Rev. D}\ }\textbf {\bibinfo {volume} {110}},\ \bibinfo
  {pages} {123533} (\bibinfo {year} {2024}{\natexlab{a}})},\ \Eprint
  {https://arxiv.org/abs/2405.00502} {arXiv:2405.00502 [astro-ph.CO]}
  \BibitemShut {NoStop}%
\bibitem [{\citenamefont {Park}\ \emph
  {et~al.}(2024{\natexlab{b}})\citenamefont {Park}, \citenamefont
  {de~Cruz~P\'erez},\ and\ \citenamefont {Ratra}}]{Park:2024pew}%
  \BibitemOpen
  \bibfield  {author} {\bibinfo {author} {\bibfnamefont {C.-G.}\ \bibnamefont
  {Park}}, \bibinfo {author} {\bibfnamefont {J.}~\bibnamefont
  {de~Cruz~P\'erez}},\ and\ \bibinfo {author} {\bibfnamefont {B.}~\bibnamefont
  {Ratra}},\ }\bibfield  {title} {\bibinfo {title} {{Is the $w_0w_a$CDM
  cosmological parameterization evidence for dark energy dynamics partially
  caused by the excess smoothing of Planck CMB anisotropy data?}},\ }\href@noop
  {} {\  (\bibinfo {year} {2024}{\natexlab{b}})},\ \Eprint
  {https://arxiv.org/abs/2410.13627} {arXiv:2410.13627 [astro-ph.CO]}
  \BibitemShut {NoStop}%
\bibitem [{\citenamefont {Park}\ and\ \citenamefont
  {Ratra}(2025)}]{Park:2025azv}%
  \BibitemOpen
  \bibfield  {author} {\bibinfo {author} {\bibfnamefont {C.-G.}\ \bibnamefont
  {Park}}\ and\ \bibinfo {author} {\bibfnamefont {B.}~\bibnamefont {Ratra}},\
  }\bibfield  {title} {\bibinfo {title} {{Is excess smoothing of Planck CMB
  ansiotropy data partially responsible for evidence for dark energy dynamics
  in other $w(z)$CDM parametrizations?}},\ }\href@noop {} {\  (\bibinfo {year}
  {2025})},\ \Eprint {https://arxiv.org/abs/2501.03480} {arXiv:2501.03480
  [astro-ph.CO]} \BibitemShut {NoStop}%
\bibitem [{\citenamefont {Ormondroyd}\ \emph {et~al.}(2025)\citenamefont
  {Ormondroyd}, \citenamefont {Handley}, \citenamefont {Hobson},\ and\
  \citenamefont {Lasenby}}]{Ormondroyd:2025iaf}%
  \BibitemOpen
  \bibfield  {author} {\bibinfo {author} {\bibfnamefont {A.~N.}\ \bibnamefont
  {Ormondroyd}}, \bibinfo {author} {\bibfnamefont {W.~J.}\ \bibnamefont
  {Handley}}, \bibinfo {author} {\bibfnamefont {M.~P.}\ \bibnamefont
  {Hobson}},\ and\ \bibinfo {author} {\bibfnamefont {A.~N.}\ \bibnamefont
  {Lasenby}},\ }\bibfield  {title} {\bibinfo {title} {{Comparison of dynamical
  dark energy with \ensuremath{\Lambda}CDM in light of DESI DR2}},\ }\href@noop
  {} {\  (\bibinfo {year} {2025})},\ \Eprint {https://arxiv.org/abs/2503.17342}
  {arXiv:2503.17342 [astro-ph.CO]} \BibitemShut {NoStop}%
\bibitem [{\citenamefont {Yang}\ \emph {et~al.}(2025)\citenamefont {Yang},
  \citenamefont {Wang}, \citenamefont {Ren}, \citenamefont {Saridakis},\ and\
  \citenamefont {Cai}}]{Yang:2025mws}%
  \BibitemOpen
  \bibfield  {author} {\bibinfo {author} {\bibfnamefont {Y.}~\bibnamefont
  {Yang}}, \bibinfo {author} {\bibfnamefont {Q.}~\bibnamefont {Wang}}, \bibinfo
  {author} {\bibfnamefont {X.}~\bibnamefont {Ren}}, \bibinfo {author}
  {\bibfnamefont {E.~N.}\ \bibnamefont {Saridakis}},\ and\ \bibinfo {author}
  {\bibfnamefont {Y.-F.}\ \bibnamefont {Cai}},\ }\bibfield  {title} {\bibinfo
  {title} {{Modified gravity realizations of quintom dark energy after DESI
  DR2}},\ }\href@noop {} {\  (\bibinfo {year} {2025})},\ \Eprint
  {https://arxiv.org/abs/2504.06784} {arXiv:2504.06784 [astro-ph.CO]}
  \BibitemShut {NoStop}%
\bibitem [{\citenamefont {Li}\ \emph {et~al.}(2025{\natexlab{a}})\citenamefont
  {Li}, \citenamefont {Wang}, \citenamefont {Zhang}, \citenamefont
  {Saridakis},\ and\ \citenamefont {Cai}}]{Li:2025cxn}%
  \BibitemOpen
  \bibfield  {author} {\bibinfo {author} {\bibfnamefont {C.}~\bibnamefont
  {Li}}, \bibinfo {author} {\bibfnamefont {J.}~\bibnamefont {Wang}}, \bibinfo
  {author} {\bibfnamefont {D.}~\bibnamefont {Zhang}}, \bibinfo {author}
  {\bibfnamefont {E.~N.}\ \bibnamefont {Saridakis}},\ and\ \bibinfo {author}
  {\bibfnamefont {Y.-F.}\ \bibnamefont {Cai}},\ }\bibfield  {title} {\bibinfo
  {title} {{Quantum Gravity Meets DESI: Dynamical Dark Energy in Light of the
  Trans-Planckian Censorship Conjecture}},\ }\href@noop {} {\  (\bibinfo {year}
  {2025}{\natexlab{a}})},\ \Eprint {https://arxiv.org/abs/2504.07791}
  {arXiv:2504.07791 [astro-ph.CO]} \BibitemShut {NoStop}%
\bibitem [{\citenamefont
  {Paliathanasis}(2025{\natexlab{a}})}]{Paliathanasis:2025hjw}%
  \BibitemOpen
  \bibfield  {author} {\bibinfo {author} {\bibfnamefont {A.}~\bibnamefont
  {Paliathanasis}},\ }\bibfield  {title} {\bibinfo {title} {{Testing
  Non-Coincident $f(Q)$-gravity with DESI DR2 BAO and GRBs}},\ }\href@noop {}
  {\  (\bibinfo {year} {2025}{\natexlab{a}})},\ \Eprint
  {https://arxiv.org/abs/2504.11132} {arXiv:2504.11132 [gr-qc]} \BibitemShut
  {NoStop}%
\bibitem [{\citenamefont {Tyagi}\ \emph {et~al.}(2025)\citenamefont {Tyagi},
  \citenamefont {Haridasu},\ and\ \citenamefont {Basak}}]{Tyagi:2025zov}%
  \BibitemOpen
  \bibfield  {author} {\bibinfo {author} {\bibfnamefont {U.~K.}\ \bibnamefont
  {Tyagi}}, \bibinfo {author} {\bibfnamefont {S.}~\bibnamefont {Haridasu}},\
  and\ \bibinfo {author} {\bibfnamefont {S.}~\bibnamefont {Basak}},\ }\bibfield
   {title} {\bibinfo {title} {{Constraints on Generalized Gravity-Thermodynamic
  Cosmology from DESI DR2}},\ }\href@noop {} {\  (\bibinfo {year} {2025})},\
  \Eprint {https://arxiv.org/abs/2504.11308} {arXiv:2504.11308 [astro-ph.CO]}
  \BibitemShut {NoStop}%
\bibitem [{\citenamefont {You}\ \emph {et~al.}(2025)\citenamefont {You},
  \citenamefont {Wang},\ and\ \citenamefont {Yang}}]{You:2025uon}%
  \BibitemOpen
  \bibfield  {author} {\bibinfo {author} {\bibfnamefont {C.}~\bibnamefont
  {You}}, \bibinfo {author} {\bibfnamefont {D.}~\bibnamefont {Wang}},\ and\
  \bibinfo {author} {\bibfnamefont {T.}~\bibnamefont {Yang}},\ }\bibfield
  {title} {\bibinfo {title} {{Dynamical Dark Energy Implies a Coupled Dark
  Sector: Insights from DESI DR2 via a Data-Driven Approach}},\ }\href@noop {}
  {\  (\bibinfo {year} {2025})},\ \Eprint {https://arxiv.org/abs/2504.00985}
  {arXiv:2504.00985 [astro-ph.CO]} \BibitemShut {NoStop}%
\bibitem [{\citenamefont {Santos}\ \emph {et~al.}(2025)\citenamefont {Santos},
  \citenamefont {Morais}, \citenamefont {Pan}, \citenamefont {Yang},\ and\
  \citenamefont {Di~Valentino}}]{Santos:2025wiv}%
  \BibitemOpen
  \bibfield  {author} {\bibinfo {author} {\bibfnamefont {F.~B. M.~d.}\
  \bibnamefont {Santos}}, \bibinfo {author} {\bibfnamefont {J.}~\bibnamefont
  {Morais}}, \bibinfo {author} {\bibfnamefont {S.}~\bibnamefont {Pan}},
  \bibinfo {author} {\bibfnamefont {W.}~\bibnamefont {Yang}},\ and\ \bibinfo
  {author} {\bibfnamefont {E.}~\bibnamefont {Di~Valentino}},\ }\bibfield
  {title} {\bibinfo {title} {{A New Window on Dynamical Dark Energy: Combining
  DESI-DR2 BAO with future Gravitational Wave Observations}},\ }\href@noop {}
  {\  (\bibinfo {year} {2025})},\ \Eprint {https://arxiv.org/abs/2504.04646}
  {arXiv:2504.04646 [astro-ph.CO]} \BibitemShut {NoStop}%
\bibitem [{\citenamefont {Alfano}\ and\ \citenamefont
  {Luongo}(2025)}]{Alfano:2025gie}%
  \BibitemOpen
  \bibfield  {author} {\bibinfo {author} {\bibfnamefont {A.~C.}\ \bibnamefont
  {Alfano}}\ and\ \bibinfo {author} {\bibfnamefont {O.}~\bibnamefont
  {Luongo}},\ }\bibfield  {title} {\bibinfo {title} {{Cosmic distance duality
  after DESI 2024 data release and dark energy evolution}},\ }\href@noop {} {\
  (\bibinfo {year} {2025})},\ \Eprint {https://arxiv.org/abs/2501.15233}
  {arXiv:2501.15233 [astro-ph.CO]} \BibitemShut {NoStop}%
\bibitem [{\citenamefont {Chaussidon}\ \emph {et~al.}(2025)\citenamefont
  {Chaussidon} \emph {et~al.}}]{Chaussidon:2025npr}%
  \BibitemOpen
  \bibfield  {author} {\bibinfo {author} {\bibfnamefont {E.}~\bibnamefont
  {Chaussidon}} \emph {et~al.},\ }\bibfield  {title} {\bibinfo {title} {{Early
  time solution as an alternative to the late time evolving dark energy with
  DESI DR2 BAO}},\ }\href@noop {} {\  (\bibinfo {year} {2025})},\ \Eprint
  {https://arxiv.org/abs/2503.24343} {arXiv:2503.24343 [astro-ph.CO]}
  \BibitemShut {NoStop}%
\bibitem [{\citenamefont {Anchordoqui}\ \emph {et~al.}(2025)\citenamefont
  {Anchordoqui}, \citenamefont {Antoniadis},\ and\ \citenamefont
  {Lust}}]{Anchordoqui:2025fgz}%
  \BibitemOpen
  \bibfield  {author} {\bibinfo {author} {\bibfnamefont {L.~A.}\ \bibnamefont
  {Anchordoqui}}, \bibinfo {author} {\bibfnamefont {I.}~\bibnamefont
  {Antoniadis}},\ and\ \bibinfo {author} {\bibfnamefont {D.}~\bibnamefont
  {Lust}},\ }\bibfield  {title} {\bibinfo {title} {{S-dual Quintessence, the
  Swampland, and the DESI DR2 Results}},\ }\href@noop {} {\  (\bibinfo {year}
  {2025})},\ \Eprint {https://arxiv.org/abs/2503.19428} {arXiv:2503.19428
  [hep-th]} \BibitemShut {NoStop}%
\bibitem [{\citenamefont {Ye}\ and\ \citenamefont {Cai}(2025)}]{Ye:2025ulq}%
  \BibitemOpen
  \bibfield  {author} {\bibinfo {author} {\bibfnamefont {G.}~\bibnamefont
  {Ye}}\ and\ \bibinfo {author} {\bibfnamefont {Y.}~\bibnamefont {Cai}},\
  }\bibfield  {title} {\bibinfo {title} {{NEC violation and ''beyond
  Horndeski'' physics in light of DESI DR2}},\ }\href@noop {} {\  (\bibinfo
  {year} {2025})},\ \Eprint {https://arxiv.org/abs/2503.22515}
  {arXiv:2503.22515 [gr-qc]} \BibitemShut {NoStop}%
\bibitem [{\citenamefont {Silva}\ \emph {et~al.}(2025)\citenamefont {Silva},
  \citenamefont {Sabogal}, \citenamefont {Souza}, \citenamefont {Nunes},
  \citenamefont {Di~Valentino},\ and\ \citenamefont {Kumar}}]{Silva:2025hxw}%
  \BibitemOpen
  \bibfield  {author} {\bibinfo {author} {\bibfnamefont {E.}~\bibnamefont
  {Silva}}, \bibinfo {author} {\bibfnamefont {M.~A.}\ \bibnamefont {Sabogal}},
  \bibinfo {author} {\bibfnamefont {M.~S.}\ \bibnamefont {Souza}}, \bibinfo
  {author} {\bibfnamefont {R.~C.}\ \bibnamefont {Nunes}}, \bibinfo {author}
  {\bibfnamefont {E.}~\bibnamefont {Di~Valentino}},\ and\ \bibinfo {author}
  {\bibfnamefont {S.}~\bibnamefont {Kumar}},\ }\bibfield  {title} {\bibinfo
  {title} {{New Constraints on Interacting Dark Energy from DESI DR2 BAO
  Observations}},\ }\href@noop {} {\  (\bibinfo {year} {2025})},\ \Eprint
  {https://arxiv.org/abs/2503.23225} {arXiv:2503.23225 [astro-ph.CO]}
  \BibitemShut {NoStop}%
\bibitem [{\citenamefont {Wolf}\ \emph {et~al.}(2025)\citenamefont {Wolf},
  \citenamefont {Garc\'\i{}a-Garc\'\i{}a}, \citenamefont {Anton},\ and\
  \citenamefont {Ferreira}}]{Wolf:2025jed}%
  \BibitemOpen
  \bibfield  {author} {\bibinfo {author} {\bibfnamefont {W.~J.}\ \bibnamefont
  {Wolf}}, \bibinfo {author} {\bibfnamefont {C.}~\bibnamefont
  {Garc\'\i{}a-Garc\'\i{}a}}, \bibinfo {author} {\bibfnamefont
  {T.}~\bibnamefont {Anton}},\ and\ \bibinfo {author} {\bibfnamefont {P.~G.}\
  \bibnamefont {Ferreira}},\ }\bibfield  {title} {\bibinfo {title} {{The
  Cosmological Evidence for Non-Minimal Coupling}},\ }\href@noop {} {\
  (\bibinfo {year} {2025})},\ \Eprint {https://arxiv.org/abs/2504.07679}
  {arXiv:2504.07679 [astro-ph.CO]} \BibitemShut {NoStop}%
\bibitem [{\citenamefont
  {Paliathanasis}(2025{\natexlab{b}})}]{Paliathanasis:2025dcr}%
  \BibitemOpen
  \bibfield  {author} {\bibinfo {author} {\bibfnamefont {A.}~\bibnamefont
  {Paliathanasis}},\ }\bibfield  {title} {\bibinfo {title} {{Dark Energy within
  the Generalized Uncertainty Principle in Light of DESI DR2}},\ }\href@noop {}
  {\  (\bibinfo {year} {2025}{\natexlab{b}})},\ \Eprint
  {https://arxiv.org/abs/2503.20896} {arXiv:2503.20896 [astro-ph.CO]}
  \BibitemShut {NoStop}%
\bibitem [{\citenamefont
  {Paliathanasis}(2025{\natexlab{c}})}]{Paliathanasis:2025cuc}%
  \BibitemOpen
  \bibfield  {author} {\bibinfo {author} {\bibfnamefont {A.}~\bibnamefont
  {Paliathanasis}},\ }\bibfield  {title} {\bibinfo {title} {{Observational
  Constraints on Dark Energy Models with $\Lambda$ as an Equilibrium Point}},\
  }\href@noop {} {\  (\bibinfo {year} {2025}{\natexlab{c}})},\ \Eprint
  {https://arxiv.org/abs/2502.16221} {arXiv:2502.16221 [astro-ph.CO]}
  \BibitemShut {NoStop}%
\bibitem [{\citenamefont {Nagpal}\ \emph {et~al.}(2025)\citenamefont {Nagpal},
  \citenamefont {Chaudhary}, \citenamefont {Gupta},\ and\ \citenamefont
  {Pacif}}]{Nagpal:2025omq}%
  \BibitemOpen
  \bibfield  {author} {\bibinfo {author} {\bibfnamefont {R.}~\bibnamefont
  {Nagpal}}, \bibinfo {author} {\bibfnamefont {H.}~\bibnamefont {Chaudhary}},
  \bibinfo {author} {\bibfnamefont {H.}~\bibnamefont {Gupta}},\ and\ \bibinfo
  {author} {\bibfnamefont {S.~K.~J.}\ \bibnamefont {Pacif}},\ }\bibfield
  {title} {\bibinfo {title} {{Late-time constraints on dynamical dark energy
  models using DESI DR2, Type Ia supernova, and CC measurements}},\ }\href
  {https://doi.org/10.1016/j.jheap.2025.100396} {\bibfield  {journal} {\bibinfo
   {journal} {JHEAp}\ }\textbf {\bibinfo {volume} {47}},\ \bibinfo {pages}
  {100396} (\bibinfo {year} {2025})}\BibitemShut {NoStop}%
\bibitem [{\citenamefont {Chen}\ and\ \citenamefont
  {Zaldarriaga}(2025)}]{Chen:2025mlf}%
  \BibitemOpen
  \bibfield  {author} {\bibinfo {author} {\bibfnamefont {S.-F.}\ \bibnamefont
  {Chen}}\ and\ \bibinfo {author} {\bibfnamefont {M.}~\bibnamefont
  {Zaldarriaga}},\ }\bibfield  {title} {\bibinfo {title} {{It's All ${\tt Ok}$:
  Curvature in Light of BAO from DESI DR2}},\ }\href@noop {} {\  (\bibinfo
  {year} {2025})},\ \Eprint {https://arxiv.org/abs/2505.00659}
  {arXiv:2505.00659 [astro-ph.CO]} \BibitemShut {NoStop}%
\bibitem [{\citenamefont {Wang}\ and\ \citenamefont
  {Mota}(2025)}]{Wang:2025bkk}%
  \BibitemOpen
  \bibfield  {author} {\bibinfo {author} {\bibfnamefont {D.}~\bibnamefont
  {Wang}}\ and\ \bibinfo {author} {\bibfnamefont {D.}~\bibnamefont {Mota}},\
  }\bibfield  {title} {\bibinfo {title} {{Did DESI DR2 truly reveal dynamical
  dark energy?}},\ }\href@noop {} {\  (\bibinfo {year} {2025})},\ \Eprint
  {https://arxiv.org/abs/2504.15222} {arXiv:2504.15222 [astro-ph.CO]}
  \BibitemShut {NoStop}%
\bibitem [{\citenamefont {Kumar}\ \emph {et~al.}(2025)\citenamefont {Kumar},
  \citenamefont {Ajith},\ and\ \citenamefont {Verma}}]{Kumar:2025etf}%
  \BibitemOpen
  \bibfield  {author} {\bibinfo {author} {\bibfnamefont {U.}~\bibnamefont
  {Kumar}}, \bibinfo {author} {\bibfnamefont {A.}~\bibnamefont {Ajith}},\ and\
  \bibinfo {author} {\bibfnamefont {A.}~\bibnamefont {Verma}},\ }\bibfield
  {title} {\bibinfo {title} {{Evidence for non-cold dark matter from DESI DR2
  measurements}},\ }\href@noop {} {\  (\bibinfo {year} {2025})},\ \Eprint
  {https://arxiv.org/abs/2504.14419} {arXiv:2504.14419 [astro-ph.CO]}
  \BibitemShut {NoStop}%
\bibitem [{\citenamefont {Luciano}\ \emph {et~al.}(2025)\citenamefont
  {Luciano}, \citenamefont {Paliathanasis},\ and\ \citenamefont
  {Saridakis}}]{Luciano:2025hjn}%
  \BibitemOpen
  \bibfield  {author} {\bibinfo {author} {\bibfnamefont {G.~G.}\ \bibnamefont
  {Luciano}}, \bibinfo {author} {\bibfnamefont {A.}~\bibnamefont
  {Paliathanasis}},\ and\ \bibinfo {author} {\bibfnamefont {E.~N.}\
  \bibnamefont {Saridakis}},\ }\bibfield  {title} {\bibinfo {title} {{Barrow
  and Tsallis entropies after the DESI DR2 BAO data}},\ }\href@noop {} {\
  (\bibinfo {year} {2025})},\ \Eprint {https://arxiv.org/abs/2504.12205}
  {arXiv:2504.12205 [gr-qc]} \BibitemShut {NoStop}%
\bibitem [{\citenamefont {Dinda}\ \emph {et~al.}(2025)\citenamefont {Dinda},
  \citenamefont {Maartens}, \citenamefont {Saito},\ and\ \citenamefont
  {Clarkson}}]{Dinda:2025svh}%
  \BibitemOpen
  \bibfield  {author} {\bibinfo {author} {\bibfnamefont {B.~R.}\ \bibnamefont
  {Dinda}}, \bibinfo {author} {\bibfnamefont {R.}~\bibnamefont {Maartens}},
  \bibinfo {author} {\bibfnamefont {S.}~\bibnamefont {Saito}},\ and\ \bibinfo
  {author} {\bibfnamefont {C.}~\bibnamefont {Clarkson}},\ }\bibfield  {title}
  {\bibinfo {title} {{Improved null tests of $\Lambda$CDM and FLRW in light of
  DESI DR2}},\ }\href@noop {} {\  (\bibinfo {year} {2025})},\ \Eprint
  {https://arxiv.org/abs/2504.09681} {arXiv:2504.09681 [astro-ph.CO]}
  \BibitemShut {NoStop}%
\bibitem [{\citenamefont
  {Choudhury}(2025)}]{choudhury2025cosmologyextendedparameterspace}%
  \BibitemOpen
  \bibfield  {author} {\bibinfo {author} {\bibfnamefont {S.~R.}\ \bibnamefont
  {Choudhury}},\ }\href {https://arxiv.org/abs/2504.15340} {\bibinfo {title}
  {Cosmology in extended parameter space with desi dr2 bao: A 2$\sigma$+
  detection of non-zero neutrino masses with an update on dynamical dark energy
  and lensing anomaly}} (\bibinfo {year} {2025}),\ \Eprint
  {https://arxiv.org/abs/2504.15340} {arXiv:2504.15340 [astro-ph.CO]}
  \BibitemShut {NoStop}%
\bibitem [{\citenamefont {Wang}\ \emph {et~al.}(2024)\citenamefont {Wang},
  \citenamefont {Abdalla}, \citenamefont {Atrio-Barandela},\ and\ \citenamefont
  {Pav\'on}}]{Wang:2024vmw}%
  \BibitemOpen
  \bibfield  {author} {\bibinfo {author} {\bibfnamefont {B.}~\bibnamefont
  {Wang}}, \bibinfo {author} {\bibfnamefont {E.}~\bibnamefont {Abdalla}},
  \bibinfo {author} {\bibfnamefont {F.}~\bibnamefont {Atrio-Barandela}},\ and\
  \bibinfo {author} {\bibfnamefont {D.}~\bibnamefont {Pav\'on}},\ }\bibfield
  {title} {\bibinfo {title} {{Further understanding the interaction between
  dark energy and dark matter: current status and future directions}},\ }\href
  {https://doi.org/10.1088/1361-6633/ad2527} {\bibfield  {journal} {\bibinfo
  {journal} {Rept. Prog. Phys.}\ }\textbf {\bibinfo {volume} {87}},\ \bibinfo
  {pages} {036901} (\bibinfo {year} {2024})},\ \Eprint
  {https://arxiv.org/abs/2402.00819} {arXiv:2402.00819 [astro-ph.CO]}
  \BibitemShut {NoStop}%
\bibitem [{\citenamefont {Amendola}(2000)}]{Amendola_2000}%
  \BibitemOpen
  \bibfield  {author} {\bibinfo {author} {\bibfnamefont {L.}~\bibnamefont
  {Amendola}},\ }\bibfield  {title} {\bibinfo {title} {Coupled quintessence},\
  }\bibfield  {journal} {\bibinfo  {journal} {Physical Review D}\ }\textbf
  {\bibinfo {volume} {62}},\ \href {https://doi.org/10.1103/physrevd.62.043511}
  {10.1103/physrevd.62.043511} (\bibinfo {year} {2000})\BibitemShut {NoStop}%
\bibitem [{\citenamefont {Zimdahl}\ \emph {et~al.}(2001)\citenamefont
  {Zimdahl}, \citenamefont {Pavón},\ and\ \citenamefont
  {Chimento}}]{Zimdahl_2001}%
  \BibitemOpen
  \bibfield  {author} {\bibinfo {author} {\bibfnamefont {W.}~\bibnamefont
  {Zimdahl}}, \bibinfo {author} {\bibfnamefont {D.}~\bibnamefont {Pavón}},\
  and\ \bibinfo {author} {\bibfnamefont {L.~P.}\ \bibnamefont {Chimento}},\
  }\bibfield  {title} {\bibinfo {title} {Interacting quintessence},\ }\href
  {https://doi.org/10.1016/s0370-2693(01)01174-1} {\bibfield  {journal}
  {\bibinfo  {journal} {Physics Letters B}\ }\textbf {\bibinfo {volume}
  {521}},\ \bibinfo {pages} {133–138} (\bibinfo {year} {2001})}\BibitemShut
  {NoStop}%
\bibitem [{\citenamefont {Chimento}\ \emph {et~al.}(2003)\citenamefont
  {Chimento}, \citenamefont {Jakubi}, \citenamefont {Pavón},\ and\
  \citenamefont {Zimdahl}}]{Chimento_2003}%
  \BibitemOpen
  \bibfield  {author} {\bibinfo {author} {\bibfnamefont {L.~P.}\ \bibnamefont
  {Chimento}}, \bibinfo {author} {\bibfnamefont {A.~S.}\ \bibnamefont
  {Jakubi}}, \bibinfo {author} {\bibfnamefont {D.}~\bibnamefont {Pavón}},\
  and\ \bibinfo {author} {\bibfnamefont {W.}~\bibnamefont {Zimdahl}},\
  }\bibfield  {title} {\bibinfo {title} {Interacting quintessence solution to
  the coincidence problem},\ }\bibfield  {journal} {\bibinfo  {journal}
  {Physical Review D}\ }\textbf {\bibinfo {volume} {67}},\ \href
  {https://doi.org/10.1103/physrevd.67.083513} {10.1103/physrevd.67.083513}
  (\bibinfo {year} {2003})\BibitemShut {NoStop}%
\bibitem [{\citenamefont {Farrar}\ and\ \citenamefont
  {Peebles}(2004)}]{Farrar_2004}%
  \BibitemOpen
  \bibfield  {author} {\bibinfo {author} {\bibfnamefont {G.~R.}\ \bibnamefont
  {Farrar}}\ and\ \bibinfo {author} {\bibfnamefont {P.~J.~E.}\ \bibnamefont
  {Peebles}},\ }\bibfield  {title} {\bibinfo {title} {Interacting dark matter
  and dark energy},\ }\href {https://doi.org/10.1086/381728} {\bibfield
  {journal} {\bibinfo  {journal} {The Astrophysical Journal}\ }\textbf
  {\bibinfo {volume} {604}},\ \bibinfo {pages} {1–11} (\bibinfo {year}
  {2004})}\BibitemShut {NoStop}%
\bibitem [{\citenamefont {Wang}\ and\ \citenamefont {Meng}(2004)}]{Wang_2004}%
  \BibitemOpen
  \bibfield  {author} {\bibinfo {author} {\bibfnamefont {P.}~\bibnamefont
  {Wang}}\ and\ \bibinfo {author} {\bibfnamefont {X.-H.}\ \bibnamefont
  {Meng}},\ }\bibfield  {title} {\bibinfo {title} {Can vacuum decay in our
  universe?},\ }\href {https://doi.org/10.1088/0264-9381/22/2/003} {\bibfield
  {journal} {\bibinfo  {journal} {Classical and Quantum Gravity}\ }\textbf
  {\bibinfo {volume} {22}},\ \bibinfo {pages} {283–294} (\bibinfo {year}
  {2004})}\BibitemShut {NoStop}%
\bibitem [{\citenamefont {Olivares}\ \emph {et~al.}(2006)\citenamefont
  {Olivares}, \citenamefont {Atrio-Barandela},\ and\ \citenamefont
  {Pavón}}]{Olivares_2006}%
  \BibitemOpen
  \bibfield  {author} {\bibinfo {author} {\bibfnamefont {G.}~\bibnamefont
  {Olivares}}, \bibinfo {author} {\bibfnamefont {F.}~\bibnamefont
  {Atrio-Barandela}},\ and\ \bibinfo {author} {\bibfnamefont {D.}~\bibnamefont
  {Pavón}},\ }\bibfield  {title} {\bibinfo {title} {Matter density
  perturbations in interacting quintessence models},\ }\bibfield  {journal}
  {\bibinfo  {journal} {Physical Review D}\ }\textbf {\bibinfo {volume} {74}},\
  \href {https://doi.org/10.1103/physrevd.74.043521}
  {10.1103/physrevd.74.043521} (\bibinfo {year} {2006})\BibitemShut {NoStop}%
\bibitem [{\citenamefont {Curbelo}\ \emph {et~al.}(2006)\citenamefont
  {Curbelo}, \citenamefont {Gonzalez}, \citenamefont {Leon},\ and\
  \citenamefont {Quiros}}]{Curbelo:2005dh}%
  \BibitemOpen
  \bibfield  {author} {\bibinfo {author} {\bibfnamefont {R.}~\bibnamefont
  {Curbelo}}, \bibinfo {author} {\bibfnamefont {T.}~\bibnamefont {Gonzalez}},
  \bibinfo {author} {\bibfnamefont {G.}~\bibnamefont {Leon}},\ and\ \bibinfo
  {author} {\bibfnamefont {I.}~\bibnamefont {Quiros}},\ }\bibfield  {title}
  {\bibinfo {title} {{Interacting phantom energy and avoidance of the big rip
  singularity}},\ }\href {https://doi.org/10.1088/0264-9381/23/5/010}
  {\bibfield  {journal} {\bibinfo  {journal} {Class. Quant. Grav.}\ }\textbf
  {\bibinfo {volume} {23}},\ \bibinfo {pages} {1585} (\bibinfo {year}
  {2006})},\ \Eprint {https://arxiv.org/abs/astro-ph/0502141}
  {arXiv:astro-ph/0502141} \BibitemShut {NoStop}%
\bibitem [{\citenamefont {Abdalla}\ \emph {et~al.}(2022)\citenamefont {Abdalla}
  \emph {et~al.}}]{Abdalla:2022yfr}%
  \BibitemOpen
  \bibfield  {author} {\bibinfo {author} {\bibfnamefont {E.}~\bibnamefont
  {Abdalla}} \emph {et~al.},\ }\bibfield  {title} {\bibinfo {title} {{Cosmology
  intertwined: A review of the particle physics, astrophysics, and cosmology
  associated with the cosmological tensions and anomalies}},\ }\href
  {https://doi.org/10.1016/j.jheap.2022.04.002} {\bibfield  {journal} {\bibinfo
   {journal} {JHEAp}\ }\textbf {\bibinfo {volume} {34}},\ \bibinfo {pages} {49}
  (\bibinfo {year} {2022})},\ \Eprint {https://arxiv.org/abs/2203.06142}
  {arXiv:2203.06142 [astro-ph.CO]} \BibitemShut {NoStop}%
\bibitem [{\citenamefont {Perivolaropoulos}\ and\ \citenamefont
  {Skara}(2022)}]{Perivolaropoulos:2021jda}%
  \BibitemOpen
  \bibfield  {author} {\bibinfo {author} {\bibfnamefont {L.}~\bibnamefont
  {Perivolaropoulos}}\ and\ \bibinfo {author} {\bibfnamefont {F.}~\bibnamefont
  {Skara}},\ }\bibfield  {title} {\bibinfo {title} {{Challenges for
  \ensuremath{\Lambda}CDM: An update}},\ }\href
  {https://doi.org/10.1016/j.newar.2022.101659} {\bibfield  {journal} {\bibinfo
   {journal} {New Astron. Rev.}\ }\textbf {\bibinfo {volume} {95}},\ \bibinfo
  {pages} {101659} (\bibinfo {year} {2022})},\ \Eprint
  {https://arxiv.org/abs/2105.05208} {arXiv:2105.05208 [astro-ph.CO]}
  \BibitemShut {NoStop}%
\bibitem [{\citenamefont {Di~Valentino}\ \emph {et~al.}(2025)\citenamefont
  {Di~Valentino} \emph {et~al.}}]{CosmoVerse:2025txj}%
  \BibitemOpen
  \bibfield  {author} {\bibinfo {author} {\bibfnamefont {E.}~\bibnamefont
  {Di~Valentino}} \emph {et~al.} (\bibinfo {collaboration} {CosmoVerse}),\
  }\bibfield  {title} {\bibinfo {title} {{The CosmoVerse White Paper:
  Addressing observational tensions in cosmology with systematics and
  fundamental physics}},\ }\href@noop {} {\  (\bibinfo {year} {2025})},\
  \Eprint {https://arxiv.org/abs/2504.01669} {arXiv:2504.01669 [astro-ph.CO]}
  \BibitemShut {NoStop}%
\bibitem [{\citenamefont {Shah}\ \emph {et~al.}(2025)\citenamefont {Shah},
  \citenamefont {Mukherjee},\ and\ \citenamefont {Pal}}]{Shah:2025ayl}%
  \BibitemOpen
  \bibfield  {author} {\bibinfo {author} {\bibfnamefont {R.}~\bibnamefont
  {Shah}}, \bibinfo {author} {\bibfnamefont {P.}~\bibnamefont {Mukherjee}},\
  and\ \bibinfo {author} {\bibfnamefont {S.}~\bibnamefont {Pal}},\ }\bibfield
  {title} {\bibinfo {title} {{Interacting dark sectors in light of DESI DR2}},\
  }\href@noop {} {\  (\bibinfo {year} {2025})},\ \Eprint
  {https://arxiv.org/abs/2503.21652} {arXiv:2503.21652 [astro-ph.CO]}
  \BibitemShut {NoStop}%
\bibitem [{\citenamefont {Pan}\ \emph {et~al.}(2025)\citenamefont {Pan},
  \citenamefont {Paul}, \citenamefont {Saridakis},\ and\ \citenamefont
  {Yang}}]{Pan:2025qwy}%
  \BibitemOpen
  \bibfield  {author} {\bibinfo {author} {\bibfnamefont {S.}~\bibnamefont
  {Pan}}, \bibinfo {author} {\bibfnamefont {S.}~\bibnamefont {Paul}}, \bibinfo
  {author} {\bibfnamefont {E.~N.}\ \bibnamefont {Saridakis}},\ and\ \bibinfo
  {author} {\bibfnamefont {W.}~\bibnamefont {Yang}},\ }\bibfield  {title}
  {\bibinfo {title} {{Interacting dark energy after DESI DR2: a challenge for
  $\Lambda$CDM paradigm?}},\ }\href@noop {} {\  (\bibinfo {year} {2025})},\
  \Eprint {https://arxiv.org/abs/2504.00994} {arXiv:2504.00994 [astro-ph.CO]}
  \BibitemShut {NoStop}%
\bibitem [{\citenamefont {Li}\ \emph {et~al.}(2024)\citenamefont {Li},
  \citenamefont {Wu}, \citenamefont {Du}, \citenamefont {Jin}, \citenamefont
  {Li}, \citenamefont {Zhang},\ and\ \citenamefont {Zhang}}]{Li:2024qso}%
  \BibitemOpen
  \bibfield  {author} {\bibinfo {author} {\bibfnamefont {T.-N.}\ \bibnamefont
  {Li}}, \bibinfo {author} {\bibfnamefont {P.-J.}\ \bibnamefont {Wu}}, \bibinfo
  {author} {\bibfnamefont {G.-H.}\ \bibnamefont {Du}}, \bibinfo {author}
  {\bibfnamefont {S.-J.}\ \bibnamefont {Jin}}, \bibinfo {author} {\bibfnamefont
  {H.-L.}\ \bibnamefont {Li}}, \bibinfo {author} {\bibfnamefont {J.-F.}\
  \bibnamefont {Zhang}},\ and\ \bibinfo {author} {\bibfnamefont
  {X.}~\bibnamefont {Zhang}},\ }\bibfield  {title} {\bibinfo {title}
  {{Constraints on Interacting Dark Energy Models from the DESI Baryon Acoustic
  Oscillation and DES Supernovae Data}},\ }\href
  {https://doi.org/10.3847/1538-4357/ad87f0} {\bibfield  {journal} {\bibinfo
  {journal} {Astrophys. J.}\ }\textbf {\bibinfo {volume} {976}},\ \bibinfo
  {pages} {1} (\bibinfo {year} {2024})},\ \Eprint
  {https://arxiv.org/abs/2407.14934} {arXiv:2407.14934 [astro-ph.CO]}
  \BibitemShut {NoStop}%
\bibitem [{\citenamefont {Li}\ \emph {et~al.}(2025{\natexlab{b}})\citenamefont
  {Li}, \citenamefont {Du}, \citenamefont {Li}, \citenamefont {Wu},
  \citenamefont {Jin}, \citenamefont {Zhang},\ and\ \citenamefont
  {Zhang}}]{Li:2025owk}%
  \BibitemOpen
  \bibfield  {author} {\bibinfo {author} {\bibfnamefont {T.-N.}\ \bibnamefont
  {Li}}, \bibinfo {author} {\bibfnamefont {G.-H.}\ \bibnamefont {Du}}, \bibinfo
  {author} {\bibfnamefont {Y.-H.}\ \bibnamefont {Li}}, \bibinfo {author}
  {\bibfnamefont {P.-J.}\ \bibnamefont {Wu}}, \bibinfo {author} {\bibfnamefont
  {S.-J.}\ \bibnamefont {Jin}}, \bibinfo {author} {\bibfnamefont {J.-F.}\
  \bibnamefont {Zhang}},\ and\ \bibinfo {author} {\bibfnamefont
  {X.}~\bibnamefont {Zhang}},\ }\bibfield  {title} {\bibinfo {title} {{Probing
  the sign-changeable interaction between dark energy and dark matter with DESI
  baryon acoustic oscillations and DES supernovae data}},\ }\href@noop {} {\
  (\bibinfo {year} {2025}{\natexlab{b}})},\ \Eprint
  {https://arxiv.org/abs/2501.07361} {arXiv:2501.07361 [astro-ph.CO]}
  \BibitemShut {NoStop}%
\bibitem [{\citenamefont {Koshelev}(2011)}]{Koshelev:2010umw}%
  \BibitemOpen
  \bibfield  {author} {\bibinfo {author} {\bibfnamefont {N.~A.}\ \bibnamefont
  {Koshelev}},\ }\bibfield  {title} {\bibinfo {title} {{On the growth of
  perturbations in interacting dark energy and dark matter fluids}},\ }\href
  {https://doi.org/10.1007/s10714-010-1113-2} {\bibfield  {journal} {\bibinfo
  {journal} {Gen. Rel. Grav.}\ }\textbf {\bibinfo {volume} {43}},\ \bibinfo
  {pages} {1309} (\bibinfo {year} {2011})},\ \Eprint
  {https://arxiv.org/abs/0912.0120} {arXiv:0912.0120 [gr-qc]} \BibitemShut
  {NoStop}%
\bibitem [{\citenamefont {Paliathanasis}\ \emph {et~al.}(2025)\citenamefont
  {Paliathanasis}, \citenamefont {Duffy}, \citenamefont {Halder},\ and\
  \citenamefont {Abebe}}]{Paliathanasis:2024abl}%
  \BibitemOpen
  \bibfield  {author} {\bibinfo {author} {\bibfnamefont {A.}~\bibnamefont
  {Paliathanasis}}, \bibinfo {author} {\bibfnamefont {K.}~\bibnamefont
  {Duffy}}, \bibinfo {author} {\bibfnamefont {A.}~\bibnamefont {Halder}},\ and\
  \bibinfo {author} {\bibfnamefont {A.}~\bibnamefont {Abebe}},\ }\bibfield
  {title} {\bibinfo {title} {{Compartmentalization and coexistence in the dark
  sector of the universe}},\ }\href
  {https://doi.org/10.1016/j.dark.2024.101750} {\bibfield  {journal} {\bibinfo
  {journal} {Phys. Dark Univ.}\ }\textbf {\bibinfo {volume} {47}},\ \bibinfo
  {pages} {101750} (\bibinfo {year} {2025})},\ \Eprint
  {https://arxiv.org/abs/2409.05348} {arXiv:2409.05348 [gr-qc]} \BibitemShut
  {NoStop}%
\bibitem [{\citenamefont {van~der Westhuizen}\ \emph
  {et~al.}(2025{\natexlab{a}})\citenamefont {van~der Westhuizen}, \citenamefont
  {Abebe},\ and\ \citenamefont {Di~Valentino}}]{vanderWesthuizen:2025II}%
  \BibitemOpen
  \bibfield  {author} {\bibinfo {author} {\bibfnamefont {M.}~\bibnamefont
  {van~der Westhuizen}}, \bibinfo {author} {\bibfnamefont {A.}~\bibnamefont
  {Abebe}},\ and\ \bibinfo {author} {\bibfnamefont {E.}~\bibnamefont
  {Di~Valentino}},\ }\href@noop {} {\bibinfo {title} {{II. Non-Linear
  Interacting Dark Energy: Analytical Solutions and Theoretical Pathologies}}}
  (\bibinfo {year} {2025}{\natexlab{a}}),\ \Eprint
  {https://arxiv.org/abs/2509.04494} {arXiv:2509.04494 [gr-qc]} \BibitemShut
  {NoStop}%
\bibitem [{\citenamefont {Li}\ and\ \citenamefont {Zhang}(2014)}]{Li_2014}%
  \BibitemOpen
  \bibfield  {author} {\bibinfo {author} {\bibfnamefont {Y.-H.}\ \bibnamefont
  {Li}}\ and\ \bibinfo {author} {\bibfnamefont {X.}~\bibnamefont {Zhang}},\
  }\bibfield  {title} {\bibinfo {title} {Large-scale stable interacting dark
  energy model: Cosmological perturbations and observational constraints},\
  }\bibfield  {journal} {\bibinfo  {journal} {Physical Review D}\ }\textbf
  {\bibinfo {volume} {89}},\ \href {https://doi.org/10.1103/physrevd.89.083009}
  {10.1103/physrevd.89.083009} (\bibinfo {year} {2014})\BibitemShut {NoStop}%
\bibitem [{\citenamefont {Gavela}\ \emph {et~al.}(2009)\citenamefont {Gavela},
  \citenamefont {Hernández}, \citenamefont {Honorez}, \citenamefont {Mena},\
  and\ \citenamefont {Rigolin}}]{M.B.Gavela_2009}%
  \BibitemOpen
  \bibfield  {author} {\bibinfo {author} {\bibfnamefont {M.}~\bibnamefont
  {Gavela}}, \bibinfo {author} {\bibfnamefont {D.}~\bibnamefont {Hernández}},
  \bibinfo {author} {\bibfnamefont {L.~L.}\ \bibnamefont {Honorez}}, \bibinfo
  {author} {\bibfnamefont {O.}~\bibnamefont {Mena}},\ and\ \bibinfo {author}
  {\bibfnamefont {S.}~\bibnamefont {Rigolin}},\ }\bibfield  {title} {\bibinfo
  {title} {Dark coupling},\ }\href
  {https://doi.org/10.1088/1475-7516/2009/07/034} {\bibfield  {journal}
  {\bibinfo  {journal} {Journal of Cosmology and Astroparticle Physics}\
  }\textbf {\bibinfo {volume} {2009}}\bibinfo  {number} { (07)},\ \bibinfo
  {pages} {034}}\BibitemShut {NoStop}%
\bibitem [{\citenamefont {Honorez}\ \emph {et~al.}(2010)\citenamefont
  {Honorez}, \citenamefont {Reid}, \citenamefont {Mena}, \citenamefont
  {Verde},\ and\ \citenamefont {Jimenez}}]{Honorez_2010}%
  \BibitemOpen
\bibfield  {number} {  }\bibfield  {author} {\bibinfo {author} {\bibfnamefont
  {L.~L.}\ \bibnamefont {Honorez}}, \bibinfo {author} {\bibfnamefont {B.~A.}\
  \bibnamefont {Reid}}, \bibinfo {author} {\bibfnamefont {O.}~\bibnamefont
  {Mena}}, \bibinfo {author} {\bibfnamefont {L.}~\bibnamefont {Verde}},\ and\
  \bibinfo {author} {\bibfnamefont {R.}~\bibnamefont {Jimenez}},\ }\bibfield
  {title} {\bibinfo {title} {Coupled dark matter-dark energy in light of near
  universe observations},\ }\href
  {https://doi.org/10.1088/1475-7516/2010/09/029} {\bibfield  {journal}
  {\bibinfo  {journal} {Journal of Cosmology and Astroparticle Physics}\
  }\textbf {\bibinfo {volume} {2010}}\bibinfo  {number} { (09)},\ \bibinfo
  {pages} {029–029}}\BibitemShut {NoStop}%
\bibitem [{\citenamefont {Lucca}(2021)}]{Lucca:2021dxo}%
  \BibitemOpen
\bibfield  {number} {  }\bibfield  {author} {\bibinfo {author} {\bibfnamefont
  {M.}~\bibnamefont {Lucca}},\ }\bibfield  {title} {\bibinfo {title} {{Dark
  energy{\textendash}dark matter interactions as a solution to the S8
  tension}},\ }\href {https://doi.org/10.1016/j.dark.2021.100899} {\bibfield
  {journal} {\bibinfo  {journal} {Phys. Dark Univ.}\ }\textbf {\bibinfo
  {volume} {34}},\ \bibinfo {pages} {100899} (\bibinfo {year} {2021})},\
  \Eprint {https://arxiv.org/abs/2105.09249} {arXiv:2105.09249 [astro-ph.CO]}
  \BibitemShut {NoStop}%
\bibitem [{\citenamefont {van~der Westhuizen}\ and\ \citenamefont
  {Abebe}(2024)}]{vanderWesthuizen:2023hcl}%
  \BibitemOpen
  \bibfield  {author} {\bibinfo {author} {\bibfnamefont {M.~A.}\ \bibnamefont
  {van~der Westhuizen}}\ and\ \bibinfo {author} {\bibfnamefont
  {A.}~\bibnamefont {Abebe}},\ }\bibfield  {title} {\bibinfo {title}
  {{Interacting dark energy: clarifying the cosmological implications and
  viability conditions}},\ }\href
  {https://doi.org/10.1088/1475-7516/2024/01/048} {\bibfield  {journal}
  {\bibinfo  {journal} {JCAP}\ }\textbf {\bibinfo {volume} {01}},\ \bibinfo
  {pages} {048}},\ \Eprint {https://arxiv.org/abs/2302.11949} {arXiv:2302.11949
  [gr-qc]} \BibitemShut {NoStop}%
\bibitem [{\citenamefont {von Marttens}\ \emph {et~al.}(2019)\citenamefont {von
  Marttens}, \citenamefont {Casarini}, \citenamefont {Mota},\ and\
  \citenamefont {Zimdahl}}]{von_Marttens_2019}%
  \BibitemOpen
  \bibfield  {author} {\bibinfo {author} {\bibfnamefont {R.}~\bibnamefont {von
  Marttens}}, \bibinfo {author} {\bibfnamefont {L.}~\bibnamefont {Casarini}},
  \bibinfo {author} {\bibfnamefont {D.}~\bibnamefont {Mota}},\ and\ \bibinfo
  {author} {\bibfnamefont {W.}~\bibnamefont {Zimdahl}},\ }\bibfield  {title}
  {\bibinfo {title} {Cosmological constraints on parametrized interacting dark
  energy},\ }\href {https://doi.org/10.1016/j.dark.2018.10.007} {\bibfield
  {journal} {\bibinfo  {journal} {Physics of the Dark Universe}\ }\textbf
  {\bibinfo {volume} {23}},\ \bibinfo {pages} {100248} (\bibinfo {year}
  {2019})}\BibitemShut {NoStop}%
\bibitem [{\citenamefont {Arévalo}\ \emph {et~al.}(2012)\citenamefont
  {Arévalo}, \citenamefont {Bacalhau},\ and\ \citenamefont
  {Zimdahl}}]{Arevalo:2011hh}%
  \BibitemOpen
  \bibfield  {author} {\bibinfo {author} {\bibfnamefont {F.}~\bibnamefont
  {Arévalo}}, \bibinfo {author} {\bibfnamefont {A.~P.}\ \bibnamefont
  {Bacalhau}},\ and\ \bibinfo {author} {\bibfnamefont {W.}~\bibnamefont
  {Zimdahl}},\ }\bibfield  {title} {\bibinfo {title} {Cosmological dynamics
  with nonlinear interactions},\ }\href
  {https://doi.org/10.1088/0264-9381/29/23/235001} {\bibfield  {journal}
  {\bibinfo  {journal} {Classical and Quantum Gravity}\ }\textbf {\bibinfo
  {volume} {29}},\ \bibinfo {pages} {235001} (\bibinfo {year}
  {2012})}\BibitemShut {NoStop}%
\bibitem [{\citenamefont {Aljaf}\ \emph {et~al.}(2021)\citenamefont {Aljaf},
  \citenamefont {Gregoris},\ and\ \citenamefont {Khurshudyan}}]{Aljaf_2021}%
  \BibitemOpen
  \bibfield  {author} {\bibinfo {author} {\bibfnamefont {M.}~\bibnamefont
  {Aljaf}}, \bibinfo {author} {\bibfnamefont {D.}~\bibnamefont {Gregoris}},\
  and\ \bibinfo {author} {\bibfnamefont {M.}~\bibnamefont {Khurshudyan}},\
  }\bibfield  {title} {\bibinfo {title} {Constraints on interacting dark energy
  models through cosmic chronometers and gaussian process},\ }\bibfield
  {journal} {\bibinfo  {journal} {The European Physical Journal C}\ }\textbf
  {\bibinfo {volume} {81}},\ \href
  {https://doi.org/10.1140/epjc/s10052-021-09306-2}
  {10.1140/epjc/s10052-021-09306-2} (\bibinfo {year} {2021})\BibitemShut
  {NoStop}%
\bibitem [{\citenamefont {Yang}\ \emph {et~al.}(2023)\citenamefont {Yang},
  \citenamefont {Pan}, \citenamefont {Mena},\ and\ \citenamefont
  {Di~Valentino}}]{Yang_2023}%
  \BibitemOpen
  \bibfield  {author} {\bibinfo {author} {\bibfnamefont {W.}~\bibnamefont
  {Yang}}, \bibinfo {author} {\bibfnamefont {S.}~\bibnamefont {Pan}}, \bibinfo
  {author} {\bibfnamefont {O.}~\bibnamefont {Mena}},\ and\ \bibinfo {author}
  {\bibfnamefont {E.}~\bibnamefont {Di~Valentino}},\ }\bibfield  {title}
  {\bibinfo {title} {On the dynamics of a dark sector coupling},\ }\href
  {https://doi.org/10.1016/j.jheap.2023.09.001} {\bibfield  {journal} {\bibinfo
   {journal} {Journal of High Energy Astrophysics}\ }\textbf {\bibinfo {volume}
  {40}},\ \bibinfo {pages} {19–40} (\bibinfo {year} {2023})}\BibitemShut
  {NoStop}%
\bibitem [{\citenamefont {Feng}\ and\ \citenamefont {Zhang}(2016)}]{Feng_2016}%
  \BibitemOpen
  \bibfield  {author} {\bibinfo {author} {\bibfnamefont {L.}~\bibnamefont
  {Feng}}\ and\ \bibinfo {author} {\bibfnamefont {X.}~\bibnamefont {Zhang}},\
  }\bibfield  {title} {\bibinfo {title} {Revisit of the interacting holographic
  dark energy model after planck 2015},\ }\href
  {https://doi.org/10.1088/1475-7516/2016/08/072} {\bibfield  {journal}
  {\bibinfo  {journal} {Journal of Cosmology and Astroparticle Physics}\
  }\textbf {\bibinfo {volume} {2016}}\bibinfo  {number} { (08)},\ \bibinfo
  {pages} {072–072}}\BibitemShut {NoStop}%
\bibitem [{\citenamefont {Zhang}\ \emph {et~al.}(2006)\citenamefont {Zhang},
  \citenamefont {Wu},\ and\ \citenamefont {Zhang}}]{Zhang_2006}%
  \BibitemOpen
\bibfield  {number} {  }\bibfield  {author} {\bibinfo {author} {\bibfnamefont
  {X.}~\bibnamefont {Zhang}}, \bibinfo {author} {\bibfnamefont {F.-Q.}\
  \bibnamefont {Wu}},\ and\ \bibinfo {author} {\bibfnamefont {J.}~\bibnamefont
  {Zhang}},\ }\bibfield  {title} {\bibinfo {title} {New generalized chaplygin
  gas as a scheme for unification of dark energy and dark matter},\ }\href
  {https://doi.org/10.1088/1475-7516/2006/01/003} {\bibfield  {journal}
  {\bibinfo  {journal} {Journal of Cosmology and Astroparticle Physics}\
  }\textbf {\bibinfo {volume} {2006}}\bibinfo  {number} { (01)},\ \bibinfo
  {pages} {003–003}}\BibitemShut {NoStop}%
\bibitem [{\citenamefont {Carrasco}\ \emph {et~al.}(2023)\citenamefont
  {Carrasco}, \citenamefont {Rincon}, \citenamefont {Saavedra},\ and\
  \citenamefont {Videla}}]{carrasco2023discriminatinginteractingdarkenergy}%
  \BibitemOpen
\bibfield  {number} {  }\bibfield  {author} {\bibinfo {author} {\bibfnamefont
  {R.}~\bibnamefont {Carrasco}}, \bibinfo {author} {\bibfnamefont
  {A.}~\bibnamefont {Rincon}}, \bibinfo {author} {\bibfnamefont
  {J.}~\bibnamefont {Saavedra}},\ and\ \bibinfo {author} {\bibfnamefont
  {N.}~\bibnamefont {Videla}},\ }\href {https://arxiv.org/abs/2310.04324}
  {\bibinfo {title} {Discriminating interacting dark energy models using
  statefinder diagnostic}} (\bibinfo {year} {2023}),\ \Eprint
  {https://arxiv.org/abs/2310.04324} {arXiv:2310.04324 [gr-qc]} \BibitemShut
  {NoStop}%
\bibitem [{\citenamefont {van~der Westhuizen}\ \emph
  {et~al.}(2025{\natexlab{b}})\citenamefont {van~der Westhuizen}, \citenamefont
  {Abebe},\ and\ \citenamefont {Di~Valentino}}]{vanderWesthuizen:2025I}%
  \BibitemOpen
  \bibfield  {author} {\bibinfo {author} {\bibfnamefont {M.}~\bibnamefont
  {van~der Westhuizen}}, \bibinfo {author} {\bibfnamefont {A.}~\bibnamefont
  {Abebe}},\ and\ \bibinfo {author} {\bibfnamefont {E.}~\bibnamefont
  {Di~Valentino}},\ }\href@noop {} {\bibinfo {title} {{I. Linear Interacting
  Dark Energy: Analytical Solutions and Theoretical Pathologies}}} (\bibinfo
  {year} {2025}{\natexlab{b}}),\ \Eprint {https://arxiv.org/abs/2509.04495}
  {arXiv:2509.04495 [gr-qc]} \BibitemShut {NoStop}%
\bibitem [{\citenamefont {van~der Westhuizen}\ \emph
  {et~al.}(2025{\natexlab{c}})\citenamefont {van~der Westhuizen}, \citenamefont
  {Abebe},\ and\ \citenamefont {Di~Valentino}}]{vanderWesthuizen:2025III}%
  \BibitemOpen
  \bibfield  {author} {\bibinfo {author} {\bibfnamefont {M.}~\bibnamefont
  {van~der Westhuizen}}, \bibinfo {author} {\bibfnamefont {A.}~\bibnamefont
  {Abebe}},\ and\ \bibinfo {author} {\bibfnamefont {E.}~\bibnamefont
  {Di~Valentino}},\ }\href@noop {} {\bibinfo {title} {{III. Interacting Dark
  Energy: Summary of Models, Pathologies, and Constraints}}} (\bibinfo {year}
  {2025}{\natexlab{c}}),\ \Eprint {https://arxiv.org/abs/2509.04496}
  {arXiv:2509.04496 [gr-qc]} \BibitemShut {NoStop}%
\bibitem [{\citenamefont {Guedezounme}\ \emph {et~al.}(2025)\citenamefont
  {Guedezounme}, \citenamefont {Dinda},\ and\ \citenamefont
  {Maartens}}]{guedezounme2025phantomcrossingdarkinteraction}%
  \BibitemOpen
  \bibfield  {author} {\bibinfo {author} {\bibfnamefont {S.~L.}\ \bibnamefont
  {Guedezounme}}, \bibinfo {author} {\bibfnamefont {B.~R.}\ \bibnamefont
  {Dinda}},\ and\ \bibinfo {author} {\bibfnamefont {R.}~\bibnamefont
  {Maartens}},\ }\href {https://arxiv.org/abs/2507.18274} {\bibinfo {title}
  {Phantom crossing or dark interaction?}} (\bibinfo {year} {2025}),\ \Eprint
  {https://arxiv.org/abs/2507.18274} {arXiv:2507.18274 [astro-ph.CO]}
  \BibitemShut {NoStop}%
\bibitem [{\citenamefont {Brinckmann}\ and\ \citenamefont
  {Lesgourgues}(2019)}]{Brinckmann:2018cvx}%
  \BibitemOpen
  \bibfield  {author} {\bibinfo {author} {\bibfnamefont {T.}~\bibnamefont
  {Brinckmann}}\ and\ \bibinfo {author} {\bibfnamefont {J.}~\bibnamefont
  {Lesgourgues}},\ }\bibfield  {title} {\bibinfo {title} {{MontePython 3:
  boosted MCMC sampler and other features}},\ }\href
  {https://doi.org/10.1016/j.dark.2018.100260} {\bibfield  {journal} {\bibinfo
  {journal} {Phys. Dark Univ.}\ }\textbf {\bibinfo {volume} {24}},\ \bibinfo
  {pages} {100260} (\bibinfo {year} {2019})},\ \Eprint
  {https://arxiv.org/abs/1804.07261} {arXiv:1804.07261 [astro-ph.CO]}
  \BibitemShut {NoStop}%
\bibitem [{\citenamefont {Lesgourgues}(2011)}]{CLASS1}%
  \BibitemOpen
  \bibfield  {author} {\bibinfo {author} {\bibfnamefont {J.}~\bibnamefont
  {Lesgourgues}},\ }\bibfield  {title} {\bibinfo {title} {{The Cosmic Linear
  Anisotropy Solving System (CLASS) I: Overview}},\ }\bibfield  {journal}
  {\bibinfo  {journal} {arXiv e-prints}\ }\href
  {https://doi.org/10.48550/arXiv.1104.2932} {10.48550/arXiv.1104.2932}
  (\bibinfo {year} {2011}),\ \Eprint {https://arxiv.org/abs/1104.2932}
  {arXiv:1104.2932 [astro-ph.IM]} \BibitemShut {NoStop}%
\bibitem [{\citenamefont {Blas}\ \emph {et~al.}(2011)\citenamefont {Blas},
  \citenamefont {Lesgourgues},\ and\ \citenamefont {Tram}}]{CLASS2}%
  \BibitemOpen
  \bibfield  {author} {\bibinfo {author} {\bibfnamefont {D.}~\bibnamefont
  {Blas}}, \bibinfo {author} {\bibfnamefont {J.}~\bibnamefont {Lesgourgues}},\
  and\ \bibinfo {author} {\bibfnamefont {T.}~\bibnamefont {Tram}},\ }\bibfield
  {title} {\bibinfo {title} {{The Cosmic Linear Anisotropy Solving System
  (CLASS). Part II: Approximation schemes}},\ }\href
  {https://doi.org/10.1088/1475-7516/2011/07/034} {\bibfield  {journal}
  {\bibinfo  {journal} {JCAP}\ }\bibfield  {number} {\bibinfo  {number} {
  (7)}},\ }\Eprint {https://arxiv.org/abs/1104.2933} {arXiv:1104.2933
  [astro-ph.CO]} \BibitemShut {NoStop}%
\bibitem [{\citenamefont {Scolnic}\ \emph {et~al.}(2022)\citenamefont {Scolnic}
  \emph {et~al.}}]{pan}%
  \BibitemOpen
  \bibfield  {author} {\bibinfo {author} {\bibfnamefont {D.}~\bibnamefont
  {Scolnic}} \emph {et~al.},\ }\bibfield  {title} {\bibinfo {title} {{The
  Pantheon+ Analysis: The Full Data Set and Light-curve Release}},\ }\href
  {https://doi.org/10.3847/1538-4357/ac8b7a} {\bibfield  {journal} {\bibinfo
  {journal} {Astrophys. J.}\ }\textbf {\bibinfo {volume} {938}},\ \bibinfo
  {pages} {113} (\bibinfo {year} {2022})},\ \Eprint
  {https://arxiv.org/abs/2112.03863} {arXiv:2112.03863 [astro-ph.CO]}
  \BibitemShut {NoStop}%
\bibitem [{\citenamefont {Akaike}(1974)}]{AIC}%
  \BibitemOpen
  \bibfield  {author} {\bibinfo {author} {\bibfnamefont {H.}~\bibnamefont
  {Akaike}},\ }\bibfield  {title} {\bibinfo {title} {{A new look at the
  statistical model identification}},\ }\href
  {https://doi.org/10.1109/TAC.1974.1100705} {\bibfield  {journal} {\bibinfo
  {journal} {IEEE Trans. Automatic Control}\ }\textbf {\bibinfo {volume}
  {19}},\ \bibinfo {pages} {716} (\bibinfo {year} {1974})}\BibitemShut
  {NoStop}%
\bibitem [{\citenamefont {Aghanim}\ \emph {et~al.}(2020)\citenamefont {Aghanim}
  \emph {et~al.}}]{Planck:2018vyg}%
  \BibitemOpen
  \bibfield  {author} {\bibinfo {author} {\bibfnamefont {N.}~\bibnamefont
  {Aghanim}} \emph {et~al.} (\bibinfo {collaboration} {Planck}),\ }\bibfield
  {title} {\bibinfo {title} {{Planck 2018 results. VI. Cosmological
  parameters}},\ }\href {https://doi.org/10.1051/0004-6361/201833910}
  {\bibfield  {journal} {\bibinfo  {journal} {Astron. Astrophys.}\ }\textbf
  {\bibinfo {volume} {641}},\ \bibinfo {pages} {A6} (\bibinfo {year} {2020})},\
  \bibinfo {note} {[Erratum: Astron.Astrophys. 652, C4 (2021)]},\ \Eprint
  {https://arxiv.org/abs/1807.06209} {arXiv:1807.06209 [astro-ph.CO]}
  \BibitemShut {NoStop}%
\end{thebibliography}%

\end{document}